\documentclass[12pt]{article}
\usepackage{psfig}
\usepackage{mathrsfs}
\usepackage{latexsym}
\usepackage{graphicx}
\usepackage{rotating}
\usepackage{amssymb}
\usepackage{flafter}
\renewcommand{\thefootnote}{\fnsymbol{footnote}}

\def\ie{\hbox{\it i.e.}{}}
\def\eg{\hbox{\it e.g.}{}}
\def\etc{\hbox{\it etc}{}}
\def\nn{\hspace{2mm}}
\def\sss{\scriptscriptstyle}
\newcommand{\s}[1]{\ensuremath{\widetilde{#1}}}   
\def\ynu{\ensuremath{Y_\nu}}
\def\ra{\ensuremath{\rightarrow}}
\def\MeV{\mbox{\rm MeV}}
\def\GeV{\mbox{\rm GeV}}
\def\eV{\mbox{\rm eV}}
\def\TeV{\mbox{\rm TeV}}

\def\sVEV#1{\left\langle #1\right\rangle}

\def\abs#1{\left| #1\right|} 

\begin{document}
\begin{titlepage}
\hfill
\vbox{
  \halign{#\hfil  \cr
    SISSA 54/2003/EP \cr
    hep-ph/0306195 \cr}}
\vspace*{4mm}
\begin{center}
  {\Large{\bf Charged Lepton Flavor Violating Decays: \\[0.3cm] 
Leading Logarithmic Approximation \\[0.55cm]
versus Full RG Results}}
  
\vspace*{7mm}
{\ S.~T.~Petcov ${}^{a, b)}$}\footnote[1]{Also at: Institute 
of Nuclear Research and Nuclear Energy, Bulgarian Academy %
of Sciences, 1784 Sofia, %
Bulgaria}\footnote[2]{E-mail: petcov@he.sissa.it},  
{\ S.~Profumo ${}^{a, b)}$}\footnote[3]{E-mail: profumo@he.sissa.it},
{\ Y.~Takanishi ${}^{c)}$}\footnote[4]{E-mail: yasutaka@ictp.trieste.it},
{\ C.~E.~Yaguna ${}^{a)}$}\footnote[5]{E-mail: yaguna@he.sissa.it} \\

\vspace*{.2cm}

{${}^{a)}${\it Scuola Internazionale Superiore di Studi 
Avanzati, I-34014 Trieste, Italy}
\vskip .15cm
${}^{b)}${\it Istituto Nazionale di Fisica Nucleare, 
Sezione di Trieste, I-34014 Trieste, Italy}
\vskip .15cm 
${}^{c)}${\it The Abdus Salam International Centre for Theoretical 
Physics, Strada Costiera 11, I-34100 Trieste, Italy}}


\end{center}
\begin{abstract}
In the context of the minimal supersymmetric extension of the Standard
Model with the right-handed Majorana neutrinos, we study lepton flavor
violating processes including full renormalization group running
effects. We systematically compare our results with the commonly used
leading logarithmic approximation, resorting to a ``best fit'' approach 
to fix
all the high energy Yukawa matrices. We find significant deviations
in large regions of the SUSY parameter space, which we outline in
detail. We also give, within this setting, some results on the
cosmo-phenomenologically preferred stau coannihilation
region. Finally, we propose a parametrization, in terms of the SUSY
input parameters, of the common SUSY mass appearing 
in the leading log and mass insertion approximation formula for 
the charged lepton flavor violating decay rates, which fits our
full renormalization group results with high precision.
\end{abstract}
June 2003
\vskip .2cm
\end{titlepage}
\newpage
\renewcommand{\thefootnote}{\arabic{footnote}}
\setcounter{footnote}{0}
\setcounter{page}{1}
\section{Introduction}
\indent 
In the Standard Model (SM) with left-handed massive neutrinos, lepton
flavor violating (LFV) processes, such as the muon decay into electron
and photon, $\mu\to e\gamma$, or tau decay into muon and photon,
$\tau\to\mu\gamma$, are extremely suppressed: the branching ratios for
these processes are typically less than $10^{-54}$~\cite{lfvsm},
henceforth far beyond experimental reach. However, this situation
changes drastically if the SM is embedded in its (minimal)
supersymmetric extension, including heavy right-handed neutrinos,
responsible for the generation of low energy neutrino masses via
the see-saw mechanism (MSSMRN)~\cite{LFV,BK}.

The compelling evidences for neutrino mixing, coming from solar
and atmospheric neutrino, and KamLAND experiments~\cite{nuexp}, 
indicate that the
lepton charges {\em are not conserved quantities}: 
as a consequence, it is not possible to choose a
flavor and gauge interaction diagonal basis in the lepton 
sector (see, $\eg$,~\cite{BP87}).

The off-diagonal elements of the supersymmetry soft-breaking (SSB)
terms in the slepton mass matrices, 
$\left(m_{\widetilde{\sss L}}^2\right)_{ij}$, 
$\left(m_{\widetilde{e}}^2\right)_{ij}$, and
trilinear scalar couplings, $\left(A_e\right)_{ij}$ ($i\not=j$)
represent new direct sources of lepton flavor (charge) 
violation which are not
suppressed by the extremely small light neutrino masses. The resulting
LFV decay branching ratios, though highly dependent on the structure of the
theory ($\eg$, the right-handed neutrino masses, neutrino Yukawa
couplings, slepton soft SUSY breaking terms $\etc$.) are typically
in the range of sensitivity of the current and planned experiments 
searching for such decays.

A critical issue in the study of the predictions for the 
LFV decay rates in the SUSY theories incorporating
the see-saw mechanism of neutrino mass generations 
is the relation between the high energy input parameters and 
the low energy phenomena. This relation is mainly dictated by radiative
corrections encompassed in the renormalization group (RG) equations.

A common approach to RG evolution in the context of LFV is to resort
to the simple {\em leading logarithmic} approximation. In \cite{BK} 
it has been shown that, in the specific case of sequential dominance in the neutrino Yukawa couplings,
 the leading log approximation may lead to an {\em underevaluation} of the 
off-diagonal entries of the slepton mass matrix. The aim of the
present work is to show in a general setting what are the effects of taking into account
the {\em full} RG running and to provide evidence of what is the
parameter space where the leading log approximation ceases to give
a reasonable approximation to the full result. We do not make any
particular assumption about the neutrino Yukawa coupling matrix, but
instead we resort to a {\em best fit} procedure, which allows us to
sit on a (particular) viable solution, compatible with the known
parameters of the neutrino sector. We also calculate, 
within this RG approach, the absolute value of some LFV decay rates
in a cosmologically and
phenomenologically allowed parameter space region of a minimal
supergravity (mSUGRA) MSSM, namely, the one where neutralino-stau
coannihilation processes take place. Finally, we present an effective
parametrization of the common SUSY mass $m_{\sss S}$ appearing in the
commonly used approximate formula for the  
$l_j\rightarrow l_i\gamma~(m_{l_{j}}>m_{l_i})$ decay branching ration,
$l_3\equiv\tau$, $l_2\equiv\mu$,  $l_1\equiv e$, in
terms of the high energy input parameters of mSUGRA, the universal
gaugino mass $M_{1/2}$ and the common SSB scalar mass $m_0$, both
taken at the GUT scale. We demonstrate that our parametrization
allows to reproduce with high precision the exact RG result
in a large region of relevant parameter space.

The paper is organized as follows: in the next
Section, we discuss the current phenomenological constraints on the
SUSY parameter space, on neutrino masses and mixings and on LFV
processes. In Section~\ref{sec:method} we illustrate our
numerical approach and methodology, while in Section~\ref{sec:RGE} we
discuss the relevant RG equations. Our results are presented in
Section~\ref{sec:results}. Finally, Section~\ref{sec:conclusions} is
devoted to conclusions. Details
regarding the formalism used for the calculations performed in the
present study are given in Appendices~\ref{sec:appA} 
and~\ref{sec:appB}.

\section{Phenomenological Setting}
\label{sec:phenosetting}
\indent
The main ingredients entailing non negligible rates for LFV processes
in the context of the MSSMRN are {\em (1)} the structure of the SSB
terms and {\em (2)} the neutrino mixing in the lepton sector of the
theory. We devote this section to the
discussion of the SUSY breaking sector 
(Sec.~\ref{sec:susyparameterspace}), which is currently restricted 
by both cosmological and accelerator data~\cite{SUSYconstr}, and 
the problem of determination
of the high energy structure of the Yukawa couplings of the theory 
(Sec.~\ref{sec:nusector}). It is well known that low energy
data allow to reconstruct only partially the high energy couplings
of the theory. Finally, we will briefly
discuss how LFV processes arise in the context of 
MSSMRN theories (Sec.~\ref{sec:LFV}).

\subsection{SUSY Parameter Space}
\label{sec:susyparameterspace}

\indent In the MSSM, the Standard Model fields are promoted to {\em superfields} and the theory
is defined by a superpotential $W$ and by a soft SUSY
breaking Lagrangian $\mathscr{L}_{\rm soft}$. This last part
results after a formal integration over the {\em hidden sector} fields
which break SUSY. In its most general setup, $\mathscr{L}_{\rm soft}$ 
depends {\em a priori} on more than one hundred parameters,
which are, however, constrained both by theoretical arguments ($\eg$,
naturalness in the Higgs sector) and experimental data ($\eg$, absence
of flavor changing neutral currents). The specific form of
$\mathscr{L}_{\rm\sss SSB}$ relies on the mechanism which communicates
SUSY breaking from the hidden to the visible sector. If one resorts to
non-trivial assumptions about high energy physics, such as Grand
Unification or specific supergravity K{\"a}hler potentials, a further
reduction of the number of parameters can occur. One of the most
common settings is the so-called {\em minimal supergravity}
(mSUGRA)~\cite{msugra}. The auxiliary chiral field giving mass to the
gauginos is supposed to lie, in mSUGRA, in the trivial representation
of the symmetric product of two adjoints of the GUT gauge symmetry
group, hence generating a {\em universal} gaugino mass, $M_{1/2}$.
Moreover, the scalar soft breaking part of the Lagrangian depends only
on a common scalar mass $m_0$ and trilinear coupling $A_0$, as well as
on $\tan\beta$. Fixing an extra sign-ambiguity in the Higgs mass term
$\mu$ completes the parameter space of mSUGRA, which then reads
\begin{equation}
\label{eq:msugraparameterspace}
M_{1/2},\ m_0,\ A_0,\ \tan\beta,\ {\rm sign}(\mu)\nn.
\end{equation}

These parameters are restricted by accelerator and cosmological constraints.
An obvious accelerator constraint comes from negative searches
for superpartners at LEP2~\cite{LEPsearches}. In mSUGRA models,
however, the most stringent bounds come
from {\em indirect} phenomenological implications of SUSY. For
instance, the limits on the almost SM like lightest $CP$-even neutral
Higgs boson put a strong constraint at $m_h>114.1~\GeV$, which,
being $m_h(m_\chi)$ an increasing function of its arguments,
translates into a lower bound on $m_\chi$~\cite{lephiggs}. Analogously, 
the inclusive branching ratio 
${\rm BR}(b\rightarrow s \gamma)$~\cite{bsgtheo} receives 
SUSY contributions
proportional to $\tan\beta$ and inversely proportional to
$m_\chi$. 
The current experimental data~\cite{bsgexp},
combined with the SM theoretical uncertainties, give the overall
bound~\cite{Baer}
\begin{equation}
\label{eq:bsgamma}
2.16\times 10^{-4}<{\rm BR}(b\rightarrow s\gamma)<4.34\times10^{-4}\nn.
\end{equation}
The bound of Eq.~(\ref{eq:bsgamma}) translates into a lower bound for
the mass of the LSP.  We also mention the constraint coming from
the deviation  $\delta a_\mu$ of the measured muon anomalous magnetic moment from its
SM value~\cite{dam}. In this case the current
theoretical uncertainties in the SM computations, mainly due to the
evaluation of the hadronic vacuum polarization contribution, make it
difficult to draw a bound from this quantity. Therefore, this bound  is not taken as a constraint in the present work.

Since supersymmetric models with conserved $R$-parity offer a natural
($\ie$ stable, massive and weakly interacting) candidate for the
non-baryonic matter content of the Universe ({\em Dark matter}),
namely the lightest neutralino $\chi$, it is natural to require that
the relic density fraction $\Omega_\chi$ of neutralinos falls into the
cosmologically preferred range. In this respect, the recent data from
WMAP~\cite{wmap} greatly increased the accuracy of previous
determination of the Dark Matter content of the Universe, indicating
that
\begin{equation}
\displaystyle \Omega_{\rm\sss CDM}\ h^2\ =\ 0.1126^{+0.0081}_{-0.0091} \nn. 
\label{eq:omegarange}
\end{equation}
As the lower limit can be evaded under the hypothesis of the existence
of another cold dark matter component besides neutralinos, we take
here as a constraint only the 
upper bound
$\Omega_{\chi}h^2\lesssim0.13$. Imposing on the mSUGRA parameter
space (\ref{eq:msugraparameterspace}) the constraints coming both from
cosmology and from accelerator experiments  
typically reduces the viable
values of the parameters $m_0$ and $M_{1/2}$ to very narrow strips~\cite{wmaprelicdens}.

The major problem, in the framework of mSUGRA,
is that the annihilations of bino-like neutralinos is generally not
sufficiently efficient, and the resulting $\Omega_\chi$ exceeds the
previously stated upper bound. Specific relic density suppression
mechanisms are needed in order to achieve reasonable values for
$\Omega_\chi$. Exhaustive investigations of the mSUGRA parameter space
showed that only four regions of the parameter space are compatible
with Eq.~(\ref{eq:omegarange})~\cite{Baer}. $(1)$ A {\em bulk} region 
where the neutralino is 
sufficiently light and no
specific suppression mechanism is needed; this region is, however,
severely restricted by increasingly accurate accelerator 
bounds. $(2)$ A {\em coannihilation} region, where the lightest 
supersymmetric particle (LSP) is quasi degenerate with 
the next-to-LSP (NLSP). In this case, the
freeze out of the two species occurs at close cosmic temperatures, and
{\em co-}annihilations between the NLSP and the LSP can drastically
reduce the relic density of the latter. $(3)$ A {\em focus} region, at high
$m_0$, close to the region excluded by the absence of radiative
electroweak symmetry breaking, where the LSP gets a non negligible
fraction of Higgsino, thus enhancing direct annihilation
into gauge bosons. $(4)$ A {\em funnel} region, where $2m_\chi\simeq m_A$,
$m_A$ being the mass of the $CP$-odd neutral Higgs boson of the
MSSM. In this region, which opens only at large $\tan\beta$, resonance
effects suppress the relic density through direct $s$-channel pole
annihilations.

In this paper we will only exploit the coannihilation region,
since the coannihilation mechanism can occur at every 
$\tan\beta$, and also takes place in a relatively 
``stable'' SUSY parameter space region, contrary to
the ``focus'' and ``funnel'' regions. Moreover, it has been recently
shown~\cite{patternSUSY} that in supergravity models where some
relation exists between the trilinear and bilinear soft supersymmetry
breaking parameters $A_0$ and $B_0$ ($\eg$, the Polonyi model or
particular no-scale models) only the coannihilation region survives
after the cosmological and phenomenological constraints are applied.

In general, at a fixed value of $\tan\beta$, the stau coannihilation
strip lies at values of \mbox{$m_0\simeq a M_{1/2} + b$}. The 
minimum $M_{1/2}$ value reflects the lower bound on $m_\chi$, dictated by
accelerator constraints, while the maximal $M_{1/2}$ value is given by the
saturation of the bound (\ref{eq:omegarange}) at $m_\chi\simeq
m_{\sss\rm NLSP}$.

\subsection{Neutrino Sector}
\label{sec:nusector}
\indent The right-handed neutrino Majorana mass term, $M_R$, and the neutrino
Yukawa couplings, $Y_\nu$, produce an effective Majorana mass matrix (see
Sec.~\ref{sec:RGE}) for the left-handed neutrinos. This is the
well-known see-saw mechanism~\cite{seesaw}:
\begin{eqnarray}
\label{eq:seesaw}
M_{\rm eff}\!\!&\simeq&\!\! M_{\nu}^{\sf T} \, M_R^{-1} \, M_{\nu} \nonumber\\
\!\!&\simeq&\!\! v_u^2 \sin^2\beta\: Y_{\nu}^{\sf T} 
M_R^{-1} Y_{\nu} \nn,
\end{eqnarray}
where $v_u$ is vacuum expectation value (VEV) of up-type 
Higgs field,  $v_u=246/\sqrt{2}~\GeV$. In what follows
we do not consider
contributions other than that in Eq.~(\ref{eq:seesaw}) 
to the left-handed Majorana mass term. 

The matrix $M_{\rm eff}$ is diagonalized by a single unitary matrix
according to
\begin{equation}
M_{\rm eff}^\mathrm{diag}
=U^{\sf T}_{\rm eff} M_{\rm eff} U_{\rm eff}
=\mathrm{diag}\left(m_1,m_2,m_3\right) \nn.  
\end{equation}
Let us define by $U_e$ the unitary matrix which diagonalizes
$M_e^\dagger M_e$, where $M_e$ is the charged 
lepton mass matrix,
\begin{equation}
U_{e}^\dagger M_{e}^\dagger M_{e} U_{e} 
= \mathrm{diag}\left(m_e^2,m_\mu^2,m_\tau^2\right) \nn, 
\end{equation}
$M_e=Y_e v_d \cos\beta$, $Y_e$ and $v_d$ being the 
charged lepton Yukawa couplings
and the down-type Higgs VEV. Then, the MNSP neutrino 
mixing matrix has the form:
\begin{eqnarray}
\label{VPMNS}
\!\!U_{\rm \sss MNSP} \!\!\!\!&=&\!\! U_{e}^\dagger U_{\rm eff} \nonumber \\
\!\!\!\!&=&\!\!\!\!\!
\left(\begin{array}{ccc} c_{13}c_{12}&c_{13}s_{12}&s_{13}e^{-i\delta}\\
    -c_{23}s_{12}-s_{23}s_{13}c_{12}e^{i\delta}&
    c_{23}c_{12}-s_{23}s_{13}s_{12}e^{i\delta}&s_{23}c_{13}\\
    s_{23}s_{12}-c_{23}s_{13}c_{12}e^{i\delta}&
    -s_{23}c_{12}-c_{23}s_{13}s_{12}e^{i\delta}&c_{23}c_{13}\end{array}\right)
\!\mathrm{diag}(e^{i\phi},e^{i\varphi},1)\,, 
\end{eqnarray}
where we have used the standard parametrization of $U_{\rm\sss MNSP}$
and the standard notations $s_{ij}=\sin\theta_{ij}$,
$c_{ij}=\cos\theta_{ij}$. In Eq.~(\ref{VPMNS}), $\delta$ is the Dirac
and $\phi$ and $\varphi$ are the two Majorana $CP$ violation 
phases~\cite{BHP80}.

The solar, atmospheric and reactor neutrino experiments~\cite{nuexp}
have shown that~\cite{global}
\begin{eqnarray}
\label{eq:experimvalues}
&& \Delta m^2_{12}\equiv\Delta m^2_\odot\simeq7.32\times 10^{-5}~\eV^2\nn, \nn\nn\nn
\Delta m^2_{23}\equiv\Delta m^2_{\rm atm}\simeq2.6\times 10^{-3}~\eV^2\,,\nonumber\\
&&\tan^2\theta_{12}\simeq0.41\nn, \nn\nn
\tan^2\theta_{\rm 23}\simeq1.0\nn,\nn\nn\sin\theta_{13}\leq 0.2\,.
\end{eqnarray}
This information does not allow us to distinguish between the three
possible types of neutrino mass spectrum: hierarchical 
($m_3\gg m_2\gg m_1$), inverted hierarchical ($m_3\ll m_2\approx m_1$) or
quasi-degenerate ($m_3\approx m_2\approx m_1$). We will assume that
the mass spectrum of light neutrinos is hierarchical and therefore
$m_3\simeq\sqrt{\Delta m^2_{\rm atm}}$ and $m_2\simeq\sqrt{\Delta
m^2_\odot}$. Additionally, we assume that all mixing angles lie in the
interval $0<\theta_{12},\theta_{23},\theta_{13}<\pi/2$.

We summarize next all assumptions we make in order to fix the Yukawa
couplings at the GUT scale:
\begin{itemize}
\item All Yukawa coupling constants are taken to be {\em real}.
\item The right-handed neutrinos are degenerate, 
${M_R}= {\widetilde M_R}~{\bf 1}$.
\item The neutrino mixing angles are in the interval $(0,\pi/2)$.
\item The CHOOZ angle is given by  $\tan\theta_{13}=0.1$; the other 
parameters are fixed to the values given in Eq.~(\ref{eq:experimvalues})
\item The neutrino spectrum is hierarchical: 
$m_3=\sqrt{\Delta m^2_{\rm atm}}$, $m_2=\sqrt{\Delta m^2_\odot}$, 
$m_1\ll m_2$.
\end{itemize}

\subsection{Lepton Flavor Violation in the MSSMRN}
\label{sec:LFV}
\indent

The see-saw mechanism requires the existence of heavy right-handed
neutrinos as well as of the neutrino Yukawa coupling $\ynu$.
Lepton flavor violation is then unavoidable because in general ${Y}_e$
and $\ynu$ cannot be diagonalized simultaneously. In the Standard
Model with the right-handed neutrinos, LFV processes, though allowed,
are suppressed well below the sensitivity of any planned experiments.
In supersymmetric models, the situation is
different because there is a new possible source of lepton flavor
violation -- the soft SUSY breaking Lagrangian,
$\mathscr{L}_\mathrm{soft}$, whose lepton part has the form:
\begin{eqnarray}
-\mathscr{L}_\mathrm{soft}&=&(m_{\sss \s L}^2)_{ij} 
\s L_i^\dagger\s L_j+( m_{\s e}^2)_{ij}
\s e_{Ri}^*\s e_{Rj}+(m_{\s \nu}^2)_{ij}
\s \nu_{Ri}^*\s \nu_{Rj}\nonumber \\
&&+\left( (A_{e})_{ij} H_d\s e_{Ri}^*\s L_j + 
(A_{\nu})_{ij} H_u \s\nu_{Ri}^*\s L_j+ h.c.\right)\,.
\end{eqnarray}
%
Non-vanishing off-diagonal elements in the slepton mass matrix are a
source of lepton flavor violation. If they are present they must be
relatively small in order to satisfy the stringent 
experimental bounds on the
rates of LFV processes~\cite{LEPsearches}:
${\rm  BR}(\mu\ra e\gamma)<1.2\times 10^{-11}$, 
${\rm BR}(\mu\ra 3e)<1\times10^{-12}$ and 
${\rm BR}(\tau\ra\mu\gamma)< 1.1\times10^{-6}$. In mSUGRA models 
it is assumed that, at the GUT scale, the
slepton mass matrices are diagonal and universal in flavor, and that
the trilinear couplings are proportional to the Yukawa couplings:
\begin{eqnarray}
\label{eq:UBC}
&& ({m}^2_{\sss \s L})_{i j}=({m}^2_{\s e})_{i j}= ({m}^2_{\s\nu})_{i j}=
({m}^2_{\s u})_{i j}=({m}^2_{\s d})_{i j}= ({m}^2_{\s {\sss Q}})_{i j}
=\delta_{i j} m_0^2\nn,\nonumber \\
&&  m_{H_d}^2=  m_{H_u}^2=m_0^2\nn,\\
&& ({A}_\nu)_{i j} = A_0 (\ynu)_{i j}\nn,\nn({A}_e)_{i j}= A_0 (Y_e)_{i j}\nn,
\nn({A}_u)_{i j}=A_0 (Y_u)_{i j}\nn,\nn ({A}_d)_{i j}= A_0 (Y_d)_{i j}\nn.
\nonumber
\end{eqnarray}
%
However, soft-breaking terms are affected by renormalization
via Yukawa and gauge interactions, so 
the LFV in the Yukawa couplings will induce LFV in 
the slepton mass matrix at low energy. The RG for 
the left-handed slepton mass matrix is given by 
\begin{eqnarray}
\mu\frac{d}{d\mu}\left({m}_{\s L}^2\right)_{j i}&=
&\mu\frac{d}{d\mu}\left({m}_{\s L}^2\right)_{j i}
\Big|_{\sss \mathrm{MSSM}}+\frac{1}{16\pi^2}
\left[ \left({m}_{\sss\s L}^2\ynu^\dagger\ynu+
\ynu^\dagger\ynu {m}_{\sss\s L}^2\right)_{j i}\right.\nonumber\\
&&\left. + 2 \left(\ynu^\dagger {m}_{\s\nu}^2\ynu+
\s m_{H_u}^2\ynu^\dagger\ynu +
{A}_\nu^\dagger{A}_\nu\right)_{j i}\right]\nonumber \nn,
\end{eqnarray}
%
where the first term is the standard $\textrm{MSSM}$ 
term which is lepton flavor conserving, while the second one is 
the source of LFV. In the leading log approximation 
and with universal boundary conditions, Eq.~(\ref{eq:UBC}), 
the off-diagonal elements of the left-handed 
slepton mass matrix at low energy are given by
\begin{equation}
\left({m}_{\sss \s L}^2\right)_{j i}\approx -\frac{1}{8\pi^2}(3 m_0^2 + A_0^2)
\left(\ynu^\dagger\ynu\right)_{j i}\log \frac{M_{\sss\rm GUT}}{M_R} \nn.
\label{sle}
\end{equation}
These off-diagonal mass terms generate new contributions in the amplitudes
of LFV processes such as $\mu\ra e+\gamma$ 
and $\tau\ra \mu+\gamma$.

\section{Method of Numerical Computation}
\label{sec:method}
\indent 
In order to perform a calculation of the rates of the lepton flavor
violation processes, we have to use Yukawa coupling constants that
reproduce correctly the low energy fermion masses and mixings. As we
have already discussed in Sec.~\ref{sec:nusector}, we have assumed
that all Yukawa coupling constants are {\em real}, and that the
down-type quark, $Y_d$, the charged lepton, $Y_e$, and the
right-handed neutrino mass matrices, $M_R$, are diagonal in the flavor
basis at the GUT scale. This implies that up-type quark Yukawa matrix
is not diagonal and is the source of the CKM mixing matrix:
\begin{table}[!t]
\begin{displaymath}
\begin{array}{lll}
\hline\hline
& {\rm Experimental~values} & \\ \hline
m_u=2.80~\MeV & m_c=0.60~\GeV & m_t=166.71~\GeV \\
m_d=4.75~\MeV & m_s=90.50~\MeV &m_b=2.90~\GeV \\
m_e=0.511~\MeV & m_\mu=105.0~\MeV & m_\tau=1.746~\GeV \\
\hline
V_{us}=0.22 & V_{cb}=0.041 &  V_{ub}=0.0035 \\
\hline
\Delta m^2_{\odot}=7.32\times10^{-5}~\eV^2 &
\Delta m^2_{\rm atm}=2.6\times10^{-3}~\eV^2 & \\
\tan^2\theta_{\odot}=0.41 & \tan^2\theta_{\rm atm}=1.0 &
\tan^2\theta_{\rm 13}=0.01 \\
\hline\hline
\end{array}
\end{displaymath}
\caption{Conventional experimental data at $M_Z$.}
\label{tab:expvs}
\end{table}
\begin{equation}
  \label{eq:CKM}
  Y_u{}_{\hspace{-0.2mm}\big|}{}_{\tiny\textrm{{\begin{tabular}{l}
\hspace{-2.4mm} GUT \\
\hspace{-2.4mm} scale \end{tabular}}}}%
=\left(\begin{array}{*{3}{c}}
              y_u & 0 & 0 \\
              0 & y_c & 0 \\
              0 & 0 & y_t \\
              \end{array}
         \right)
\times V^{\rm 3 \times 3}_{\rm\sss CKM} \nn.
\end{equation}
In the standard parametrization the CKM matrix reads:
\begin{equation}
  \label{eq:CKMp}
 V^{\rm 3 \times 3}_{\rm\sss CKM}=
\left(\begin{array}{*{3}{c}}
\s c_{12}\s c_{13}  & \s s_{12}\s c_{13} & \s s_{13} \\
-\s s_{12}\s c_{23}-\s c_{12}\s s_{23}\s s_{13}  &
\s c_{12}\s c_{23}-\s s_{12}\s s_{23}\s s_{13} &
\s s_{23}\s c_{13} \\
\s s_{12}\s s_{23} -\s c_{12}\s c_{23}\s s_{13} &
-\s c_{12}\s s_{23}-\s s_{12}\s c_{23}\s s_{13} &
\s c_{23}\s c_{13} \\
\end{array}\right)  \nn,
\end{equation}
where $\s c_{ij}=\cos\s\theta_{ij}$ and $\s s_{ij}=\sin\s\theta_{ij}$, 
$i,j=1,2,3$ being generation indices, and we have neglected the 
$CP$ violation phase.

Let us note that the assumption that the matrix of the charged
lepton Yukawa couplings, $Y_e$, and the right-handed Majorana mass
matrix, $M_R$, can always be simultaneously diagonalized at the GUT
scale is not fulfilled in all GUT theories. Assuming that $Y_e$ and
$M_R$ are diagonal implies that the solar and atmospheric neutrino
mixing angles are generated essentially by the neutrino Yukawa
couplings, $Y_\nu$, since $Y_e$ and $M_R$ do not change substantially
by the RG effects. In other words, the MNSP matrix is essentially the
diagonalizing matrix of the see-saw light neutrino mass matrix. If the
charge lepton Yukawa matrix is not diagonal, its diagonalization will
contribute significantly to the neutrino mixing in the weak charged
lepton current.

Under the assumption made of real $Y_\nu$, the latter depends on
$9$ parameters at the GUT scale. In the numerical computation we
calculate the quark, charged lepton, the two heavier neutrino masses,
and the CKM and MNSP mixing angles, using a set of $21$ free
parameters contained in $Y_u$ (Eq.~(\ref{eq:CKM})), $Y_d$, $Y_e$,
$Y_\nu$ ($6 + 3+ 3 + 9=21$) at the GUT scale. We treat $M_R$ as an
input parameter. We use the one-loop renormalization group
equations\footnote{In the renormalization group equations for 
the gauge 
coupling constants and gaugino mass terms we have included
a part of the two-loop contributions as well. For further details
see Appendix~\ref{sec:appA}.} to obtain the values
of these parameters at the weak scale, set to the mass of the $Z$
boson ($M_Z=91.188~\GeV$). In order to find the ``best possible fit'',
we define a quantity called \mbox{\rm b.p.f.}, which depends on the
ratios of the fitted masses and mixing angles to the experimentally
determined masses and mixing angles at the energy scale of $M_Z$ (see
in Tab.~\ref{tab:expvs}):
\begin{equation}
\label{gof}
\mbox{\rm b.p.f.}\equiv\sum \left[\ln \left(
\frac{\sVEV{f}}{f_{\rm exp}} \right) \right]^2 \nn.
\end{equation}
Here $\sVEV{f}$ are the fitted masses and mixing angles with a set of
Yukawa coupling constants and $f_{\rm exp}$ are the
corresponding experimental values. We select Yukawa coupling constants
which give a minimal value of \mbox{\rm b.p.f.} less than $10^{-2}$.
In other words, the Yukawa coupling constants which we will use for the
numerical calculation of the lepton flavor 
violation decay rates can fit the $17$ low energy
fermion masses and mixing angles within an average 
deviation from the experimental values of $\exp(\sqrt{10^{-2}/17})
\approx 1.025$, $\ie$,
we can reproduce the indicated low energy values
with a deviation from the measured ones
which on average is less than $3\%$.

\section{Renormalization Group Equations from the
Universal Right-Handed Neutrino Scale}
\label{sec:RGE}
\indent

In this section we will discuss the renormalization group equations
treatment of the neutrino Yukawa couplings, the right-handed Majorana
mass term and of the trilinear coupling term, $A_\nu$. From the 
gauge coupling unification scale -- the GUT scale -- to the
universal Majorana right-handed neutrino mass scale we use the MSSM
renormalization equations which are given 
in Appendix~\ref{sec:appA}. Below the see-saw scale, $M_R$, the 
right-handed neutrinos are
{\em integrated out}. Thus, $M_R$, $Y_\nu$ and
$A_\nu$ are  not present in our set of renormalization group equations
below $M_R$, since they are no longer physically relevant: their effects 
are in fact encompassed into the running of the effective neutrino mass matrix 
$M_{\rm eff}$ below that scale. This implies that the RGE for the up-type and leptonic
Yukawa $\beta$-functions below the universal right-handed Majorana
mass scale and up to the supersymmetry breaking scale, $M_{\rm susybk}$,
\begin{equation}
  \label{eq:susybk}
  M_{\rm susybk} \!\equiv\! \sqrt{m_{\widetilde t_1} m_{\widetilde t_2}}\nn,
\end{equation}
where $m_{\widetilde t_1}$ and $m_{\widetilde t_2}$ are the stop masses,
read:
\begin{eqnarray}
  \label{eq:yumrms}
 16\pi^2 \frac{d}{d t} Y_{u_{ij}} &\!=\!&
\left \{ -\frac{13}{15} g_1^2 - 3 g_2^2 -\frac{16}{3} g_3^2
  + 3 \,{\rm Tr} ( Y_u Y_u^{\dagger} )\right \} Y_{u_{ij}}\nonumber \\
&& \; + 3 \, ( Y_u Y_u^{\dagger} Y_u)_{ij}
      + ( Y_u Y_{d}^{\dagger} Y_{d})_{ij}\nn,  \\
16\pi^2 \frac{d}{d t} Y_{e_{ij}} &\!=\!&
\left \{ -\frac{9}{5} g_1^2 - 3 g_2^2
  + 3 \,{\rm Tr} ( Y_d Y_d^{\dagger})
  +   {\rm Tr} ( Y_e Y_e^{\dagger})
\right \} Y_{e_{ij}} + 3 \, \left( Y_e Y_e^{\dagger} Y_e \right)_{ij} \nn, \\
16\pi^2 \frac{d}{d t} \left( m^2_{\widetilde{\sss L}}
\right)_{ij}&\!=\!&   -\left( \frac{6}{5} g_1^2 \left| M_1 \right|^2
+ 6 g_2^2 \left| M_2 \right|^2 \right) \delta_{ij}
-\frac{3}{5} g_1^2~S~\delta_{ij} \nonumber \\
&&
+ \left( m^2_{\widetilde{\sss L}} Y_e^{\dagger} Y_e
+ Y_e^{\dagger} Y_e m^2_{\widetilde{\sss L}}\right)_{ij}
+ 2 \left( Y_e^{\dagger} m^2_{\widetilde e} Y_e
           +{m}^2_{H_d} Y_e^{\dagger} Y_e
+ A_e^{\dagger} A_e \right)_{ij} \nn,\\
%
16\pi^2 \frac{d}{d t} A_{e_{ij}}&\!=\!&
 \left\{ -\frac{9}{5} g_1^2 -3 g_2^2
+ 3 {\rm Tr} ( Y_d^{\dagger} Y_d )
+   {\rm Tr} ( Y_e^{\dagger} Y_e ) \right \} A_{e_{ij}} \nonumber \\
&&
+ 2 \left\{
\frac{9}{5} g_1^2 M_1 + 3 g_2^2 M_2
+ 3 {\rm Tr} ( Y_d^{\dagger} A_d)
+   {\rm Tr} ( Y_e^{\dagger} A_e) \right \} Y_{e_{ij}} \nonumber \\
&&
+ 4 \left( Y_e Y_e^{\dagger} A_e \right)_{ij}
+ 5 \left( A_e Y_e^{\dagger} Y_e \right)_{ij}\nn, \\
16\pi^2 \frac{d}{d t} A_{u_{ij}}&=&
\left\{ -\frac{13}{15} g_1^2 - 3 g_2^2 - \frac{16}{3} g_3^2 + 3 {\rm Tr}
(Y_u^{\dagger} Y_u)\right \} A_{u_{ij}}
\nonumber \\
&&
+ 2 \left\{ \frac{13}{15} g_1^2 M_1 + 3 g_2^2 M_2 +\frac{16}{3} g_3^2 M_3
+ 3 {\rm Tr}(Y_u^{\dagger} A_u) \right \} Y_{u_{ij}}
\nonumber \\
&&
+ 4 ( Y_u Y_u^{\dagger} A_u)_{ij}
+ 5 ( A_u Y_u^{\dagger} Y_u)_{ij}
+ 2 ( Y_u Y_d^{\dagger} A_d)_{ij}
+ ( A_u Y_d^{\dagger} Y_d)_{ij} \nn,\\
{16 \pi^2} \frac{d}{d t} (m^2_{H_u}) &=&  -\left(\frac{6}{5}
g^{2}_{1} \abs{M_{1}}^{2} + 6 g^{2}_{2} \abs{M_{2}}^{2}\right)
+ \frac{3}{5} g_1^2 S
\nonumber\\
&&
+ 6\, {\rm {\rm Tr}}\left(
m^2_{\widetilde Q} Y^{\dagger}_{u} Y_{u} + Y^{\dagger}_{u}
( m^2_{\widetilde u} + m^2_{H_u} ) Y_{u} + A^{\dagger}_{u} A_{u}
\right) \nn.
\end{eqnarray}

Note that the other renormalization group equations do not change in
this energy interval, $\ie$, we apply the same equations for the gauge
coupling constants, gaugino mass terms, $Y_d$, 
$m^2_{\sss\widetilde  Q}$, $m^2_{\widetilde u}$, $m^2_{\widetilde d}$, 
$m^2_{\widetilde e}$, $A_d$ and $m^2_{{H_d}}$ as in Appendix~\ref{sec:appA}.

 In an analogous way, we take into account the running of the effective
neutrino mass matrix considered as a non-renormalizable term~\cite{5run}
from the universal right-handed Majorana neutrino mass scale to the
supersymmetry breaking scale, Eq.~(\ref{eq:susybk}):
\begin{equation}
  \label{eq:5dim}
 16\pi^2 \frac{d}{d t} M_{\rm eff}= \left\{ -\frac{6}{5} g_1^2 -6 g^2_2
+ 6~{\rm Tr}~(Y_u^\dagger Y_u)\right\} \, M_{\rm eff} 
+ (Y^\dagger_e Y_e)^{\sf T}
M_{\rm eff} +  M_{\rm eff} (Y^\dagger_e Y_e) \nn.
\end{equation}

Below the supersymmetry breaking scale we use the usual
renormalization group equations for the Standard Model 
and non-supersymmetric version of the see-saw mass matrix
renormalization equation (see the set of
renormalization group equations in~\cite{NT}). We use as well 
the relation between the the Weinberg-Salam Higgs field, 
$\phi_{\sss\rm WS}$, self-coupling constant 
$\lambda$ and the gauge coupling constants
\begin{equation}
  \label{eq:selfcoup}
\lambda = \frac{1}{4} \left(g_1^2 +  g_2^2 \right) \nn.
\end{equation}
This relation is a consequence of the underlying supersymmetric
structure of the theory (see, $\eg$, \cite{martin}).

\section{Results}
\label{sec:results}
\indent We present in this section the main results of this paper. We
begin discussing the corrections to the leading log approximate values
of the off-diagonal entries of the slepton mass matrix stemming from
the full RGE running. We then review the corresponding effects on LFV
processes, and discuss the r{\^o}le of the sign of the trilinear
coupling $A_0$ and of $\mu$. Finally, we give predictions for 
LFV decay rates, within the present best-fit approach, in the coannihilation
region of mSUGRA models and present an effective parametrization of
the common SUSY scale $m_{\sss S}$ appearing in the ``short-hand'' 
formula for ${\rm BR}(l_j\rightarrow l_i)$ (see Eq.~(\ref{eq:min})). 
We include in our results predictions at low, intermediate and large $\tan\beta$. 
In particular, we took $\tan\beta=3$ in order to illustrate the effects we find in the extremely low $\tan\beta$ regime, 
though in some regions of the parameter space this value is ruled out by the LEP2 bound on $m_h$.

\subsection{The Left-Handed Sleptons Mass Matrix}
\begin{figure}[!h]
\begin{center}
\begin{tabular}{cc}
\includegraphics[scale=0.5,angle=-90]{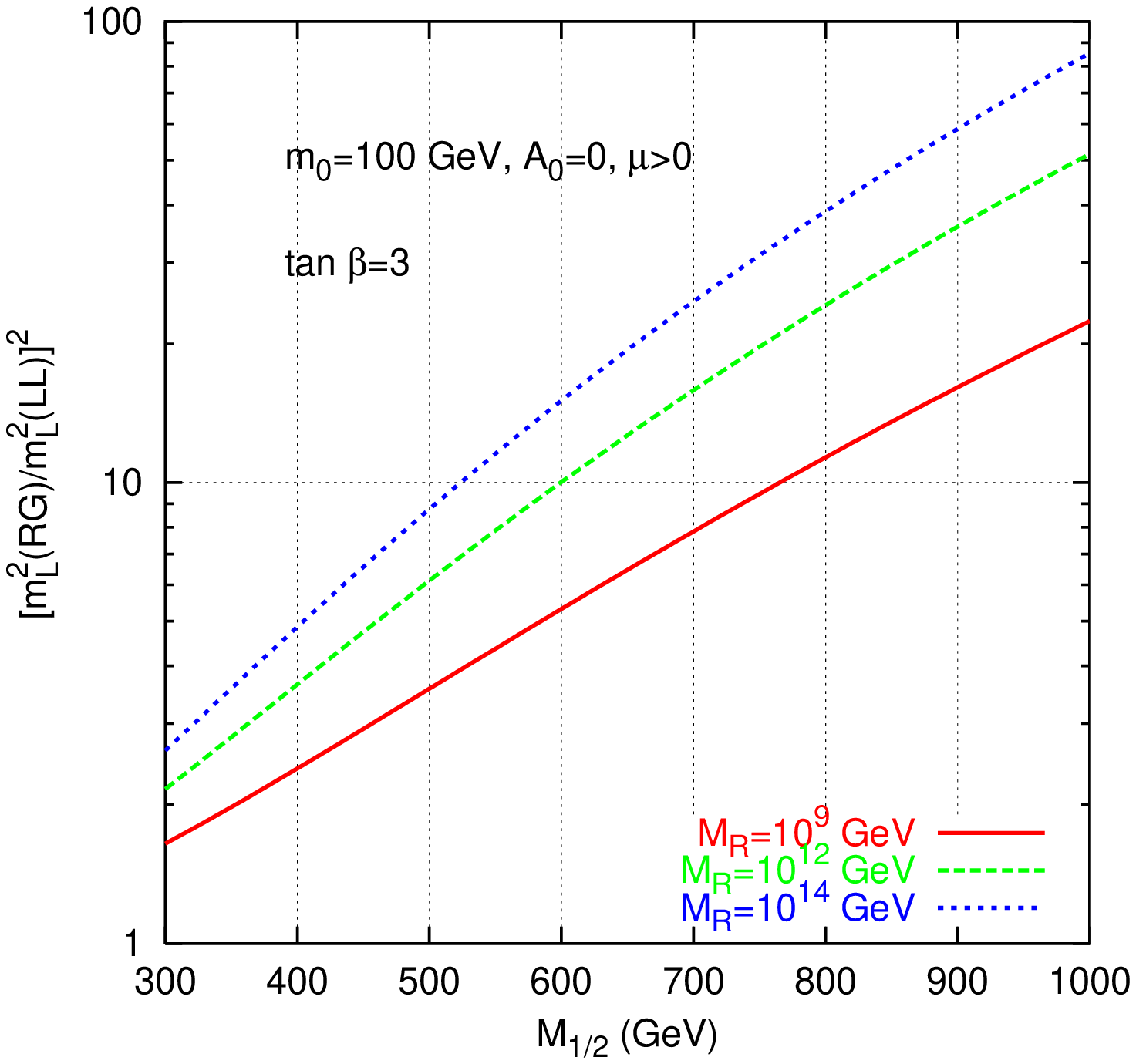} &
\includegraphics[scale=0.5,angle=-90]{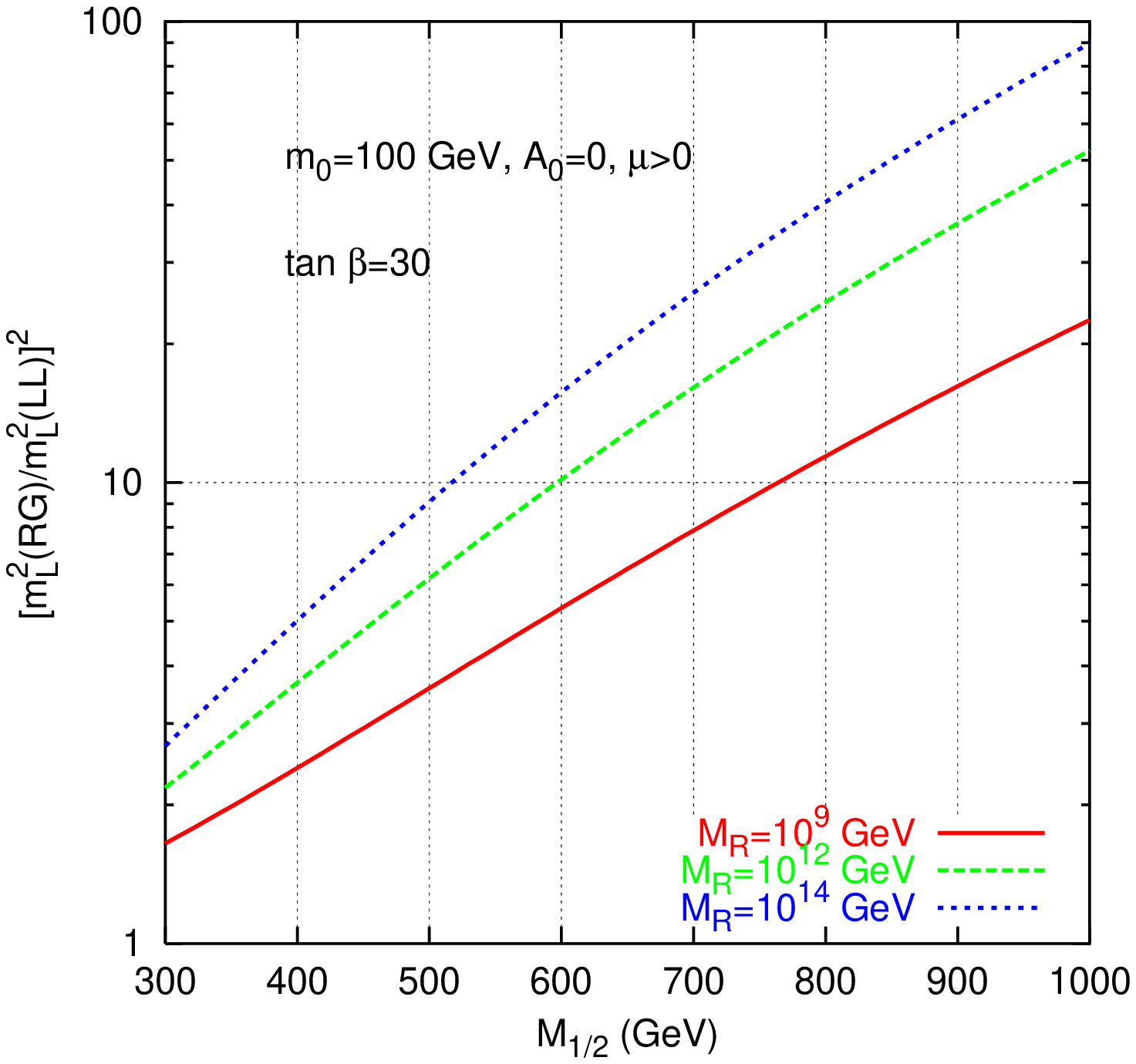}\\
&\\[-0.2cm]
\hspace{1.cm} (\emph{a}) & \hspace{1.cm} (\emph{b})\\
\end{tabular}
\caption{\em\small The ratio of the (2,1) element of 
$\left[m^2_{\widetilde L}(RG)/m^2_{\widetilde L}(LL)\right]^2$ 
as a function of $M_{1/2}$ at
$\tan\beta=3.0$ ({\em a}) and $\tan\beta=30.0$ ({\em b}). In the two
panels we fixed $m_0=100~\GeV$, $A_0=0$ and $\mu>0$.}
\label{fig:ODfixM0}
\end{center}
\end{figure}

\indent
We investigate the effect of full RG running in the off-diagonal
entries of the slepton mass matrix $m^2_{\widetilde {\sss L}}$ by
numerically evaluating the ratio of the exact result and of the
leading log approximation result which reads
\begin{equation}
\label{eq:leadinglog}
\left(m^2_{\widetilde {\sss L}}(LL)\right)_{ji}\equiv 
-\frac{1}{16\pi^2}(6m_0^2+2A_0^2)\left(Y^\dagger_\nu Y_\nu\right)_{ji}
\log\left(\frac{M_{\sss GUT}}{M_{\sss R}}\right)
\end{equation}
We study in Fig.~\ref{fig:ODfixM0} the element (2,1) of the quantity
$\left[m^2_{\widetilde {\sss L}}(RG)/m^2_{\widetilde {\sss L}}(LL)\right]^2$ 
as a function of the common right-handed
neutrino mass scale $M_R$ for two representative values of $\tan\beta=3.0,\ 
30.0$. We first notice that, in all cases, full RG running yields an
{\em increase} in the off-diagonal slepton mass matrix elements with
respect to the leading log approximation. Secondly, increasing $M_R$
gives rise to larger effects, whose size increase with the high energy
common gaugino mass $M_{1/2}$. This is expected, since the effect of
$M_{1/2}$ in the running is completely disregarded in the leading log
approximation. For $M_R$ up to $10^9~\GeV$ we get a
maximal increase of one order of magnitude at $M_{1/2}\simeq1~\TeV$,
while for $M_R\approx 10^{14}~\GeV$ the error one makes taking the
leading log approximation amounts to {\em two orders of magnitude}.
{}Finally, a remarkably weak dependence on $\tan\beta$ is found, as
is also indicated by 
the comparison of the two panels in Fig.~\ref{fig:ODfixM0}.

\begin{figure}[!h]
\begin{center}
\begin{tabular}{cc}
\includegraphics[scale=0.5,angle=-90]{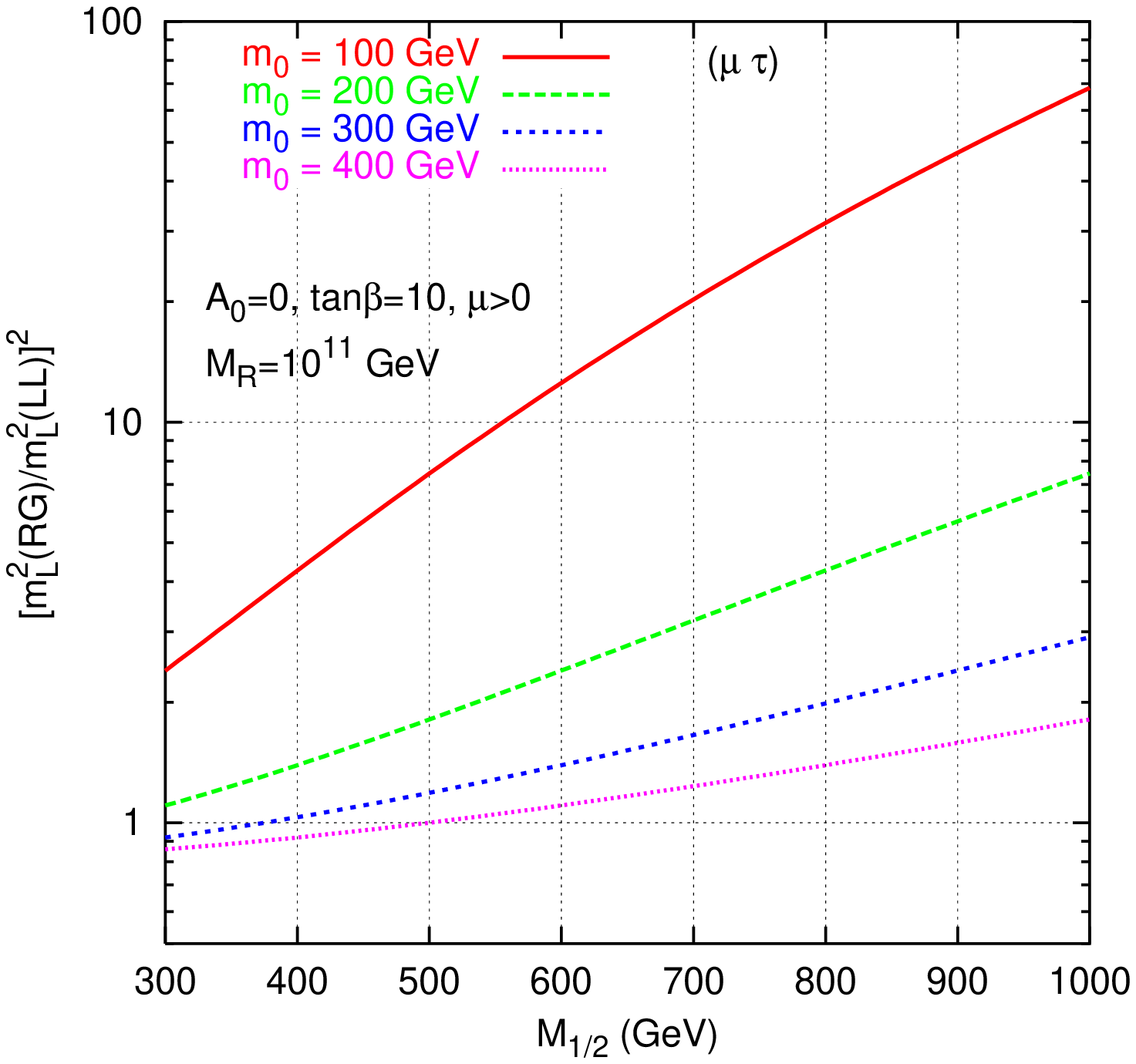} &
\includegraphics[scale=0.5,angle=-90]{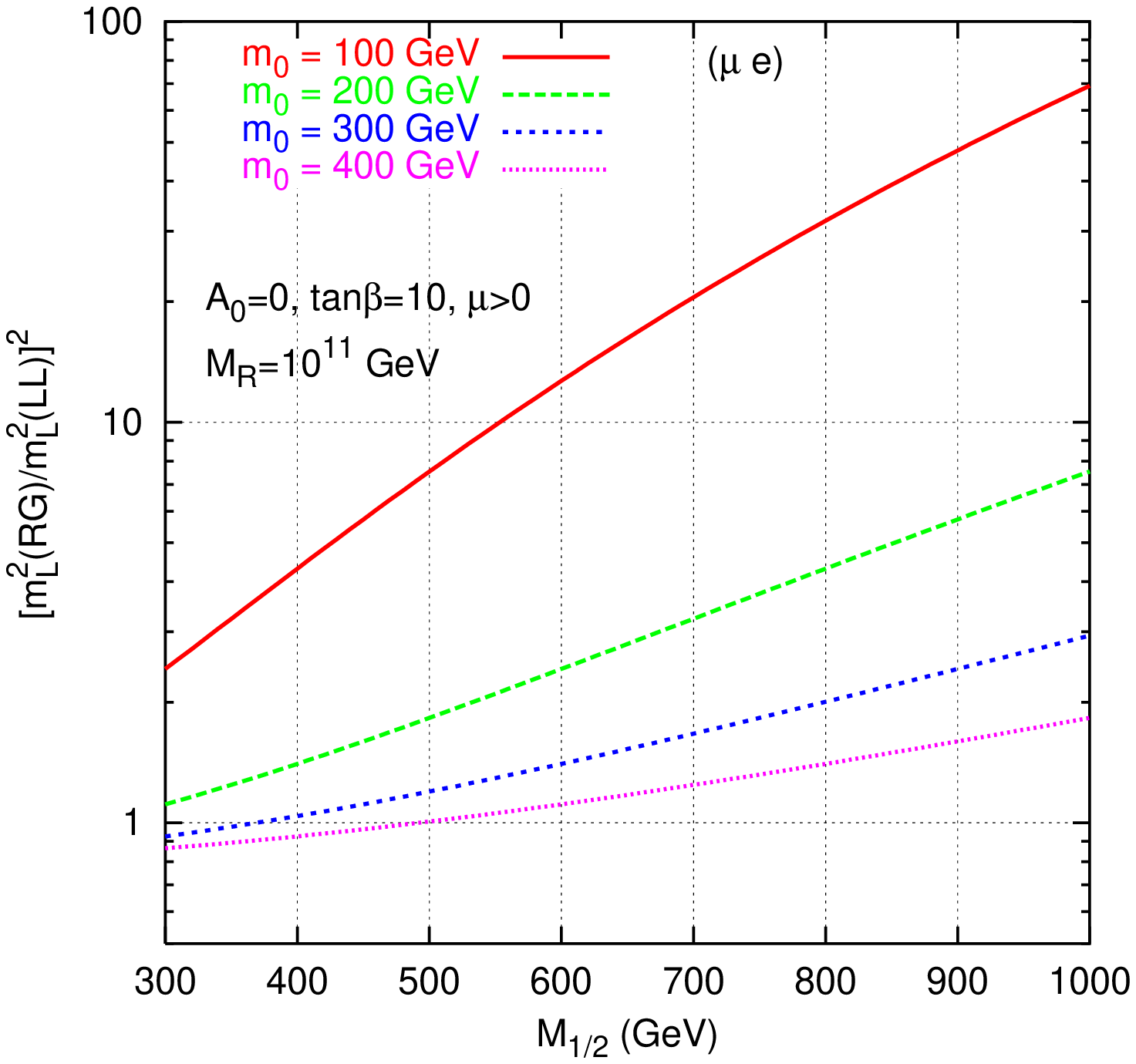}\\
&\\[-0.2cm]
\hspace{1.cm} (\emph{a}) & \hspace{1.cm} (\emph{b})\\
\end{tabular}
\caption{\em\small The ratio of the (3,2) ({\em a}) element and of 
the of (2,1) ({\em b}) element of 
$\left[m^2_{\widetilde L}(RG)/m^2_{\widetilde L}(LL)\right]^2$
as a function of $M_{1/2}$ at $\tan\beta=10.0$. In the two panels we
fixed $M_R=10^{11}~\GeV$, $A_0=0$ and $\mu>0$.}
\label{fig:ODfixMR}
\end{center}
\end{figure}

In Fig.~\ref{fig:ODfixMR} we study the $(3,2)$ and $(2,1)$ entries of the
same ratio at $\tan\beta=10.0$ at various values of $m_0$, setting
$M_R=10^{11}~\GeV$. We notice that increasing $m_0$ reduces the
effects of the full RG running: for the maximal value of 
 $m_0=400~\GeV$ we consider, 
the ratio of interest
$\left[m^2_{\widetilde {\sss L}}(RG)/m^2_{\widetilde {\sss L}}(LL)\right]^2$
changes at most by a factor of $\sim 2$.

We then conclude that the leading log approximation is not accurate
for the evaluation of the off-diagonal entries of the slepton mass
matrix in the low $m_0$ and large $M_{1/2}$ regime. The degree of
accuracy is further worsened with the 
increasing of the right-handed neutrino
mass scale $M_R$.

\subsection{Lepton Flavor Violating Processes}
\indent

In this section we turn to the main topic of interest in the present
work: the size of the corrections induced by the full RG running in
the rates of LFV processes. We compare here the results obtained
inserting the leading log approximation for
$\left(m^2_{\widetilde{\sss L}}(LL)\right)_{ji}$,
Eq.~(\ref{eq:leadinglog}), in the full mass matrix
formulae~\cite{MSSMRN} given in Appendix~\ref{sec:appB}, 
see Eqs.~(\ref{eq:amp})-(\ref{eq:BR}), 
with the exact RG running result for the LFV decay
branching ratios. We would 
like to emphasize that the results for the off-diagonal slepton
mass terms derived in leading log approximation and by 
exact RG running are included in the slepton mass matrix which after that is
diagonalized and the mass eigenstate 
scalar leptons and the corresponding
mixing matrix determined. The letter are used in the calculations of
${\rm BR}(\l_j\to l_i\gamma)$. Note that 
the branching ratios of the LFV processes
depend on the eigenvalues of the 
the slepton mass matrix
and on the elements of the matrix which
{\em diagonalize} it (see Appendix~\ref{sec:appB});
they do not depend {\em directly} on 
the off-diagonal entries of the slepton
mass matrix. Therefore we will in general get
results which differ from the ones expected
on the basis of the results 
reported in the preceding section.

\begin{figure}[!h]
\begin{center}
\begin{tabular}{cc}
\includegraphics{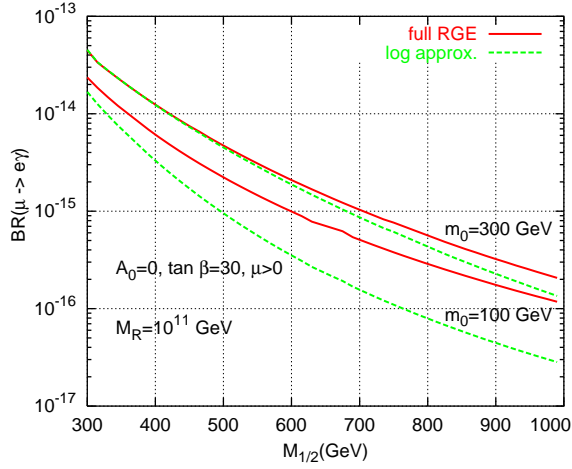} &
\includegraphics{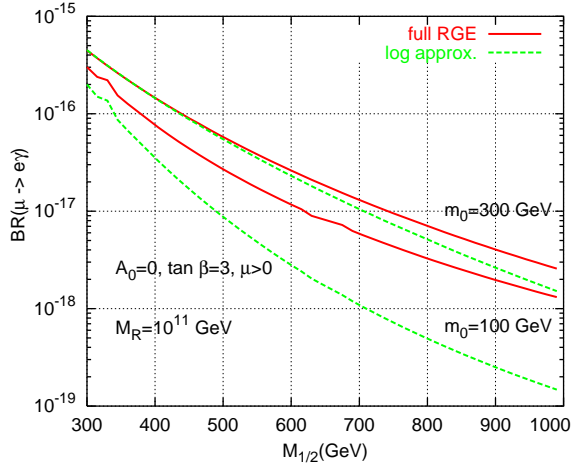}\\
&\\[-0.2cm]
\hspace{1.cm} (\emph{a}) & \hspace{1.cm} (\emph{b})\\
\end{tabular}
\caption{\em\small The branching ratio for the LFV process
${\rm BR}(\mu\rightarrow e \gamma)$ as a function of $M_{1/2}$ at 
$m_0=100~\GeV$ and $m_0=100~\GeV$ in the full RG (solid red lines)
and leading logarithmic approximation (dashed green lines). We set in
the figures $M_R=10^{11}~\GeV$, $A_0=0$, $\mu>0$ and
$\tan\beta=30.0$ ({\em a}) and $\tan\beta=3.0$ ({\em b}).}
\label{fig:hisano}
\end{center}
\end{figure}

In Fig.~\ref{fig:hisano} we plot as functions of $M_{1/2}$ the full RG
results (solid red lines) and the results obtained within the leading log
approximation (dashed green lines) for two values of $m_0$,
$m_0 =100;~300~\GeV$, at $\tan\beta=30.0$ and 
$M_R=10^{11}~\GeV$. We see that, in agreement with 
was obtained in the preceding section,
the increasing of $m_0$ reduces the RG corrections: the RG flow is
mainly driven by $m_0$ and the corrections due to $M_{1/2}$ become
less important. We get, for $m_0=100~\GeV$, a correction
of roughly one order of magnitude at large $M_{1/2}\approx1~\TeV$,
while for the $m_0=300~\GeV$ case the effect reduces to a factor of 
two. We notice, comparing panels ({\em a}) and ({\em b}), that 
the difference between the exact RG results and those obtained
in the leading log approximation is larger at lower values 
of $\tan\beta$.

\begin{figure}[!h]
\begin{center}
\begin{tabular}{cc}
\includegraphics{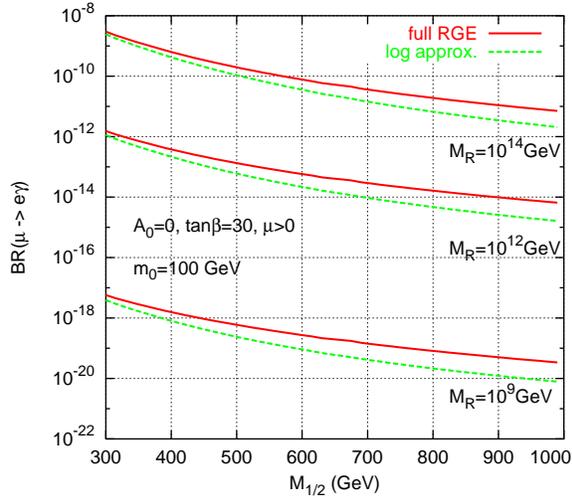} &
\includegraphics{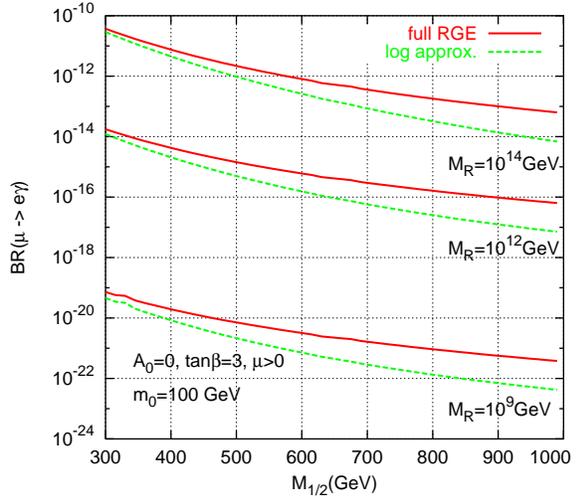}\\
&\\[-0.2cm]
\hspace{1.cm} (\emph{a}) & \hspace{1.cm} (\emph{b})\\
\end{tabular}
\caption{\em\small The branching ratio for the LFV process
${\rm BR}(\mu\rightarrow e \gamma)$ as a function of $M_{1/2}$ at
$M_R=10^{9}~\GeV,\ 10^{12}~\GeV$ and $10^{14}~\GeV$ in the full RG (solid
red lines) and leading logarithmic approximation (dashed green
lines). We set in the figures $m_0=100~\GeV$, $A_0=0$, $\mu>0$
and $\tan\beta=30.0$ (\emph{a}) and $\tan\beta=3.0$ (\emph{b}).}
\label{fig:meg}
\end{center}
\end{figure}

{}Fig.~\ref{fig:meg} shows the effect
on ${\rm BR}(\mu\to e\gamma)$ of changing the right-handed
heavy neutrino mass $M_R$ as a function of $M_{1/2}$. We used
$m_0=100~\GeV$, $A_0=0$, $\mu>0$ and $\tan\beta=30.0$, taking
$M_R=10^{9}~\GeV,\ 10^{12}~\GeV$ and $10^{14}~\GeV$. We find that the
size of the difference between the exact RG result and the leading log
approximation one depends weakly on $M_R$. At large $M_{1/2}$ we
always get a discrepancy between leading log and full RG results of
about one order of magnitude (notice the different scale with respect
to the preceding figure). Here, again, the effect is enhanced at lower
$\tan\beta$ (see panel ({\em b})).

{}Finally, we would like to emphasize that although the 
change in the LFV decay rates we find using 
the exact RG results in comparison with those
obtained in the leading log approximation 
are within one order of magnitude,
it would be mandatory to 
use the former in order to draw 
conclusions concerning the SUSY sector
if any LF violating decay 
will be observed and its branching ratio measured. 
As it follows from Figs.~\ref{fig:meg},
if we fix ${\rm BR}(\mu\rightarrow e\gamma)$, for
instance, we could make an error of up to 
$\approx 50\%$ using the leading log approximation
in deriving a {\em lower bound} on $M_{1/2}$. 


\subsection{Exact RG Evolution Effects from $\mu$ and $A_0$}
\begin{figure}[!h]
\begin{center}
\begin{tabular}{cc}
\includegraphics[scale=0.5,angle=-90]{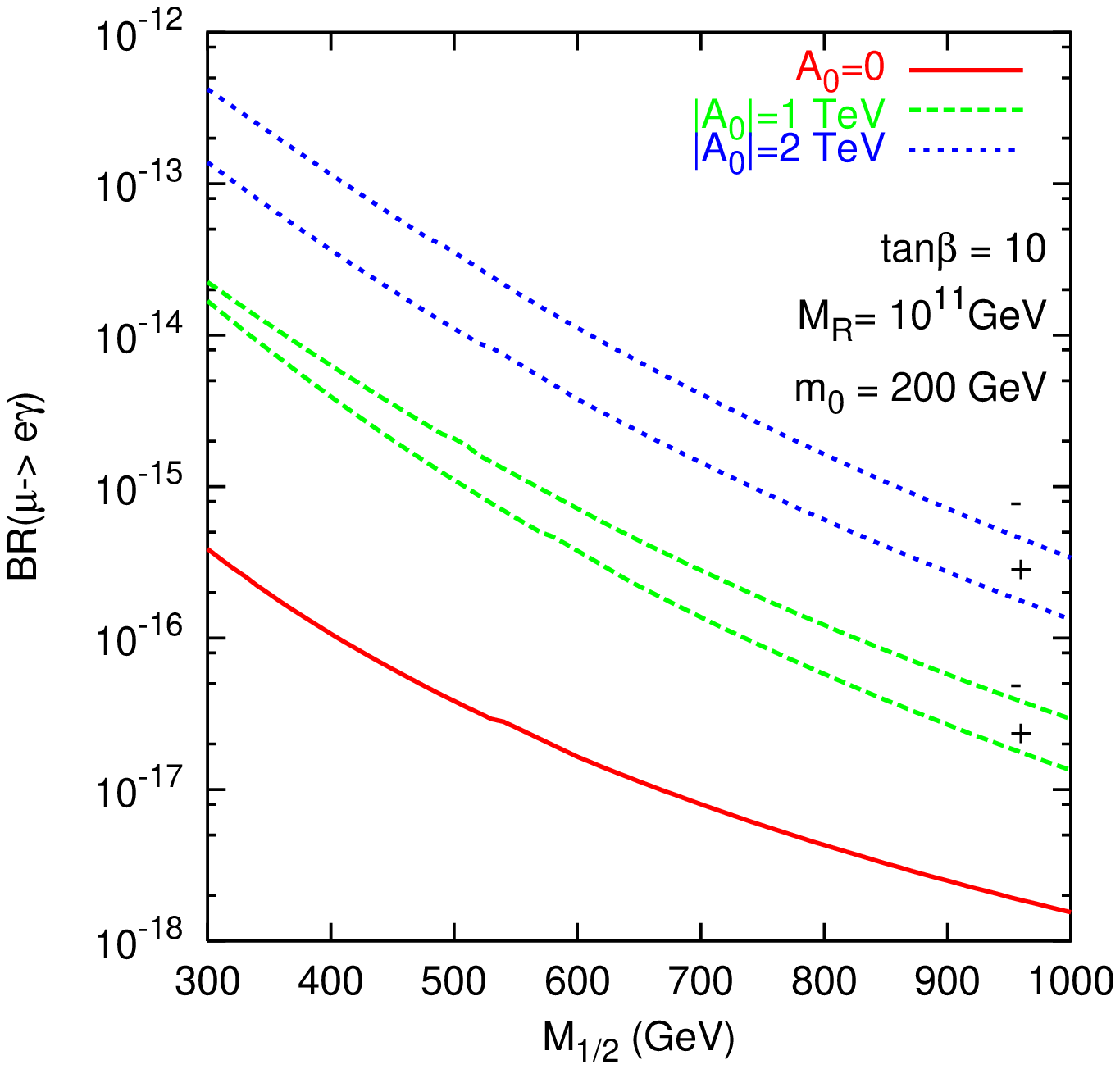} &
\includegraphics[scale=0.5,angle=-90]{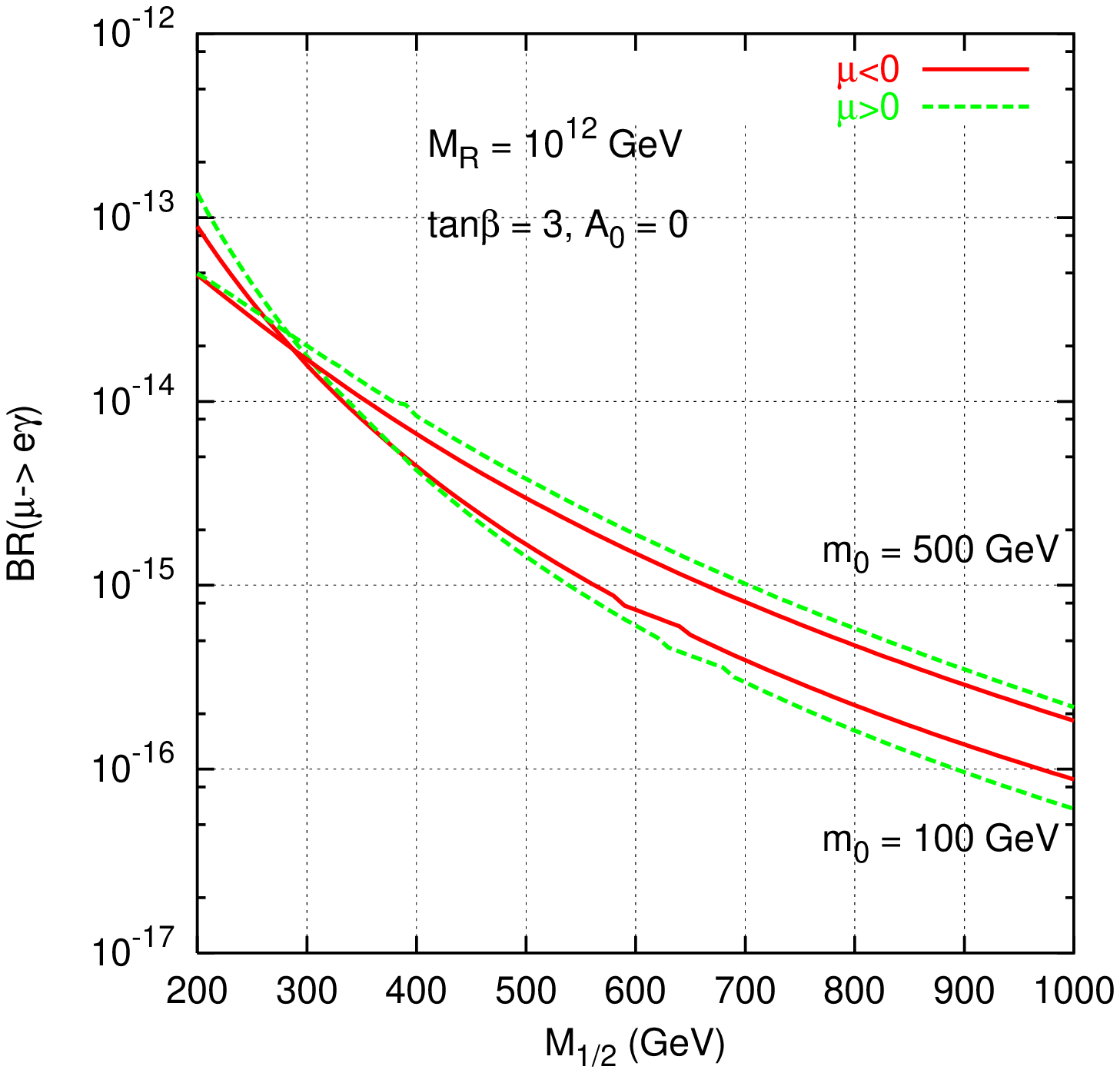}\\
&\\[-0.2cm]
\hspace{1.cm} (\emph{a}) & \hspace{1.cm} (\emph{b})\\
\end{tabular}
\caption{\em\small (\emph{a}): ${\rm BR}(\mu\rightarrow e\gamma)$ as a
function of $M_{1/2}$ at $A_0=0,\ \pm1~\TeV,\ \pm2~\TeV$
for $\tan\beta=10.0$, $M_R=10^{11}~\GeV$ and $m_0=200~\GeV$. 
(\emph{b}): The same at $m_0=100~\GeV$ (lower lines) and
$m_0=500~\GeV$ (upper lines) at $\mu>0$ (dashed green lines) and
$\mu<0$ (solid red lines), for $\tan\beta=3.0$, $M_R=10^{12}~\GeV$.}
\label{fig:amu}
\end{center}
\end{figure}

\indent
In this section we deal with two other high energy input
``parameters'' which are neglected in the leading log approximation,
namely the {\em sign} of the scalar trilinear coupling $A_0$ and of
$\mu$. We depict in the left panel of Fig.~\ref{fig:amu} the results
we get, in the full RG running computation, for $A_0=0$ (solid red
line), $A_0=\pm1~\TeV$ (resp. lower and upper dashed green lines) and
$A_0=\pm2~\TeV$ (resp. lower and upper dotted blue lines). In the
figure we fixed for definiteness $\tan\beta=10.0$, $M_R=10^{11}~\GeV$
and $m_0=200~\GeV$. We see that instead {\em negative} values of 
the trilinear
coupling generically {\em enhance}, by up to roughly a factor of $5$,
the LFV branching ratios. Flipping the sign of $\mu$ gives rise to
smaller corrections (panel ({\em b})). In particular, we find that at
low $m_0$, values of $\mu>0$ slightly {\em suppress} the LFV branching
ratios, while at larger $m_0$ a mild enhancement takes place.

\subsection{The Coannihilation region}
\indent

\begin{figure}[!h]
\begin{center}
\begin{tabular}{cc}
\includegraphics[scale=0.5,angle=-90]{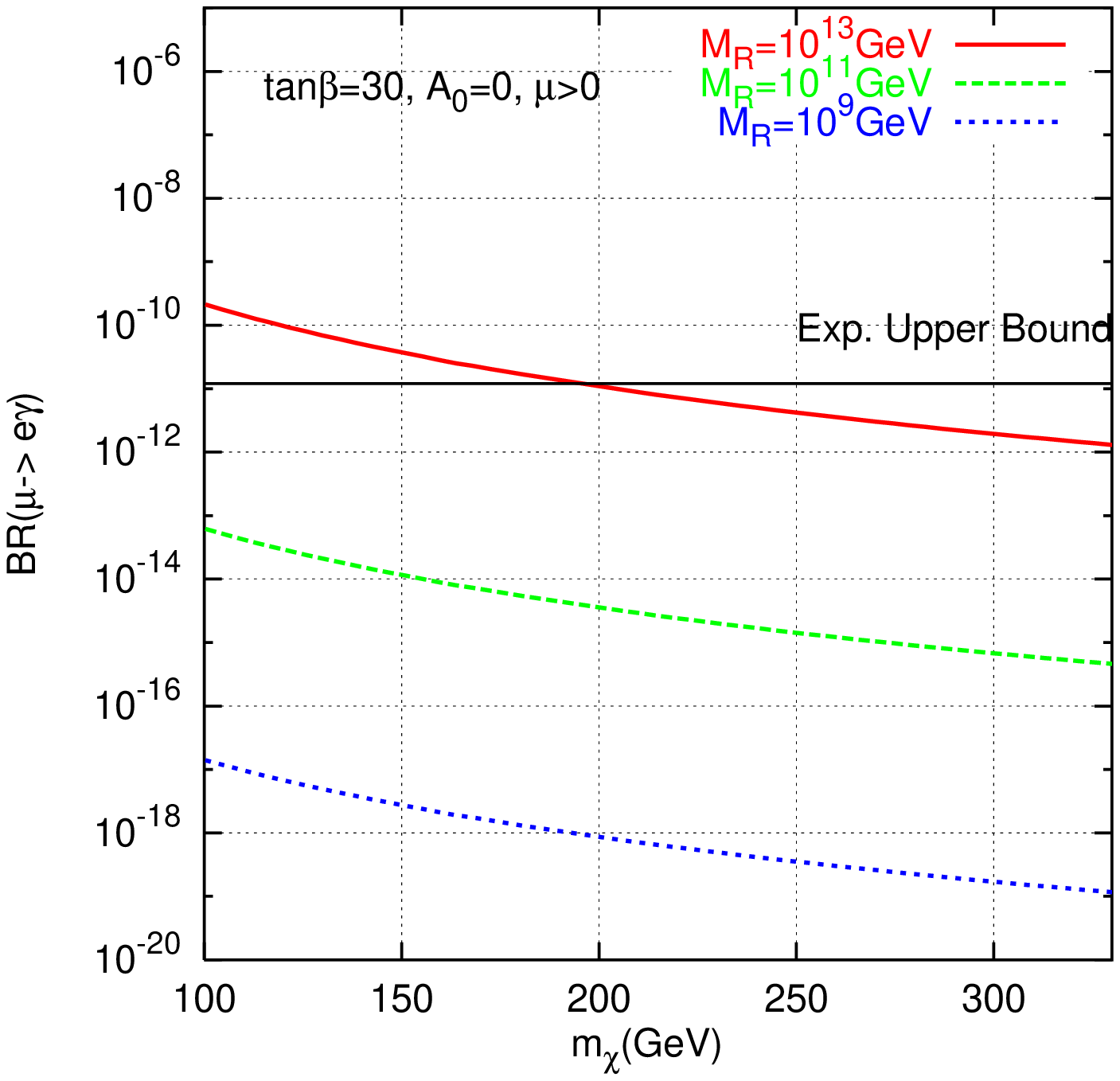} &
\includegraphics[scale=0.5,angle=-90]{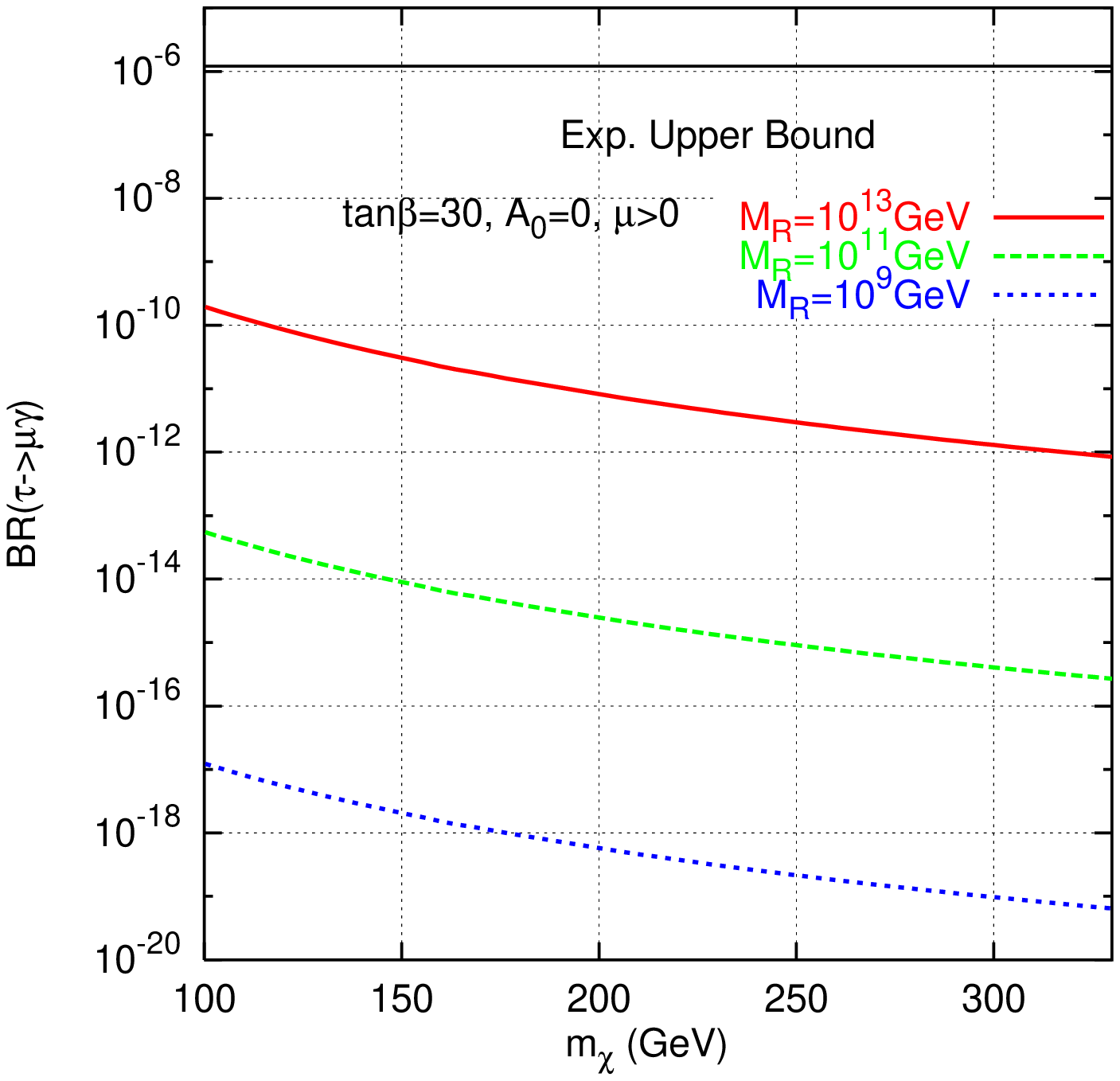}\\
&\\[-0.2cm]
\hspace{1.cm} (\emph{a}) & \hspace{1.cm} (\emph{b})\\
\end{tabular}
\caption{\em\small (\emph{a}): ${\rm BR}(\mu\rightarrow e\gamma)$ as a
function of $m_\chi$ in the stau coannihilation strip at
$\tan\beta=30.0$, $\mu>0$ and $A_0=0$, for $M_R=10^9~\GeV$, 
$10^{11}~\GeV$ and $10^{13}~\GeV$. 
(\emph{b}): the same for ${\rm BR}(\tau\rightarrow\mu\gamma)$.}
\label{fig:CoannTB30}
\end{center}
\end{figure}

\begin{figure}[!h]
\begin{center}
\begin{tabular}{cc}
\includegraphics[scale=0.5,angle=-90]{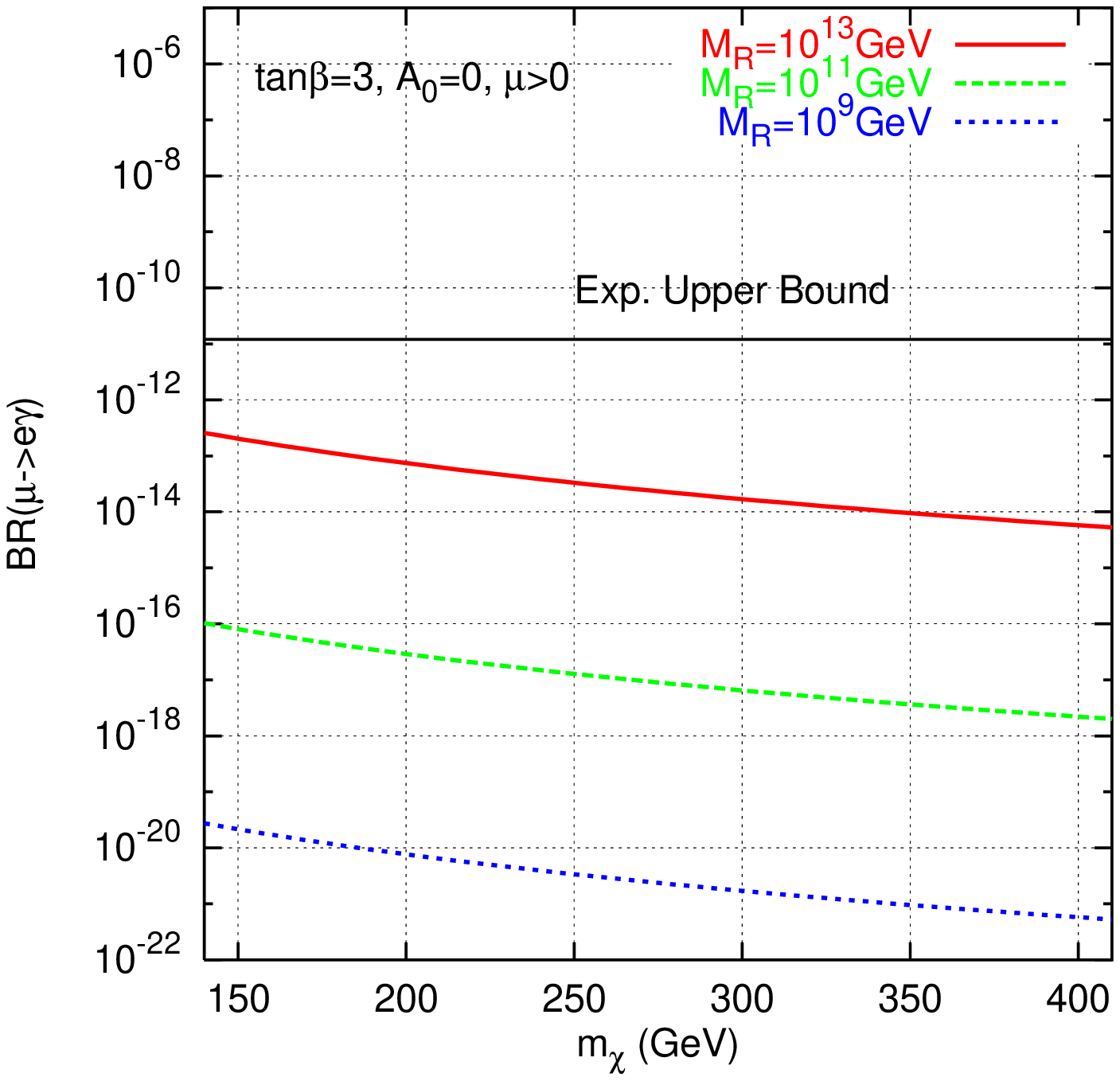} &
\includegraphics[scale=0.5,angle=-90]{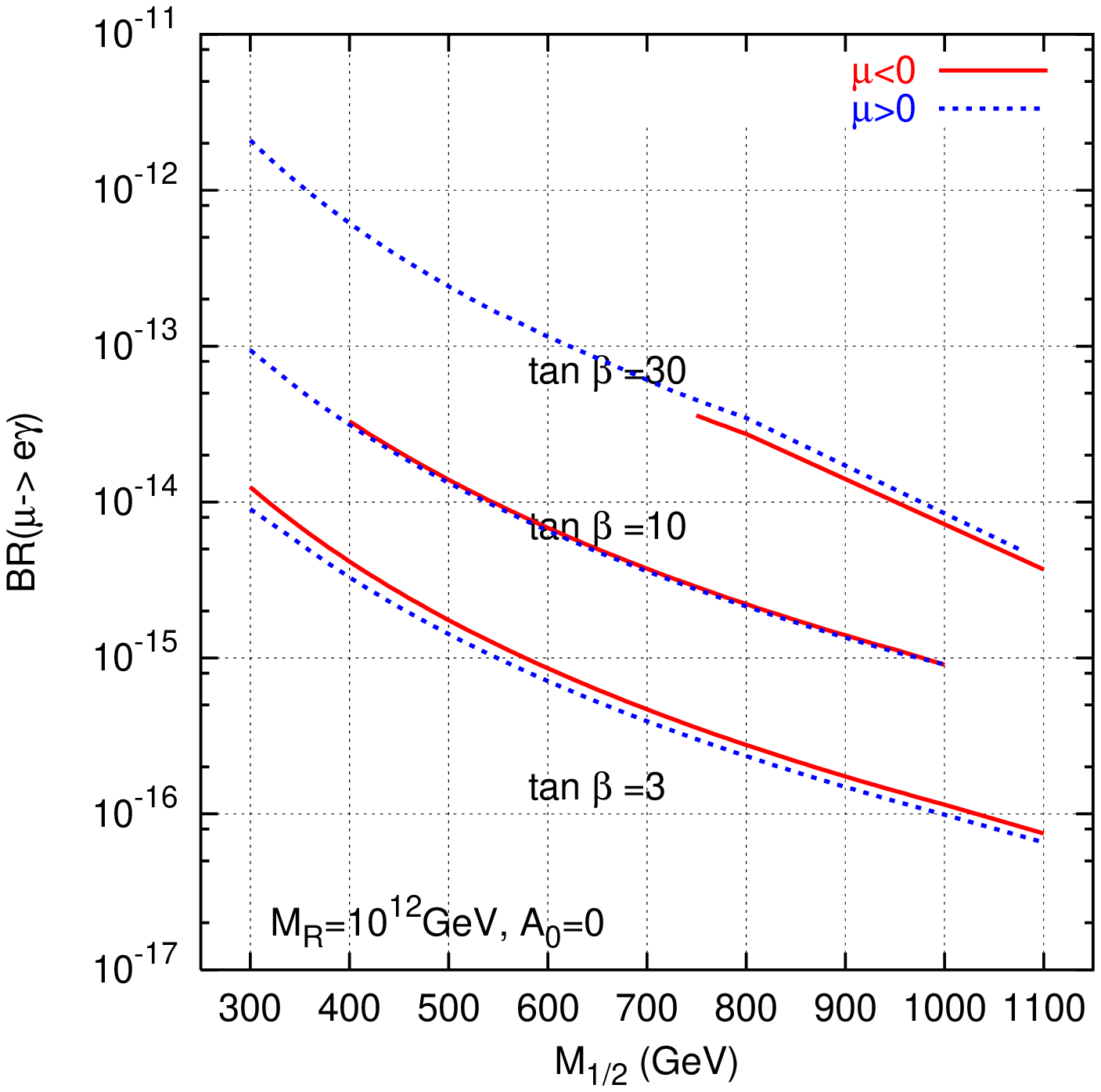}\\
&\\[-0.2cm]
\hspace{1.cm} (\emph{a}) & \hspace{1.cm} (\emph{b})\\
\end{tabular}
\caption{\em\small (\emph{a}): Same as in Fig.\ref{fig:CoannTB30}
(\emph{a}) but at $\tan\beta=3.0$. (\emph{b}) 
${\rm BR}(\mu\rightarrow e\gamma)$ along the cosmologically 
and phenomenologically viable
coannihilation strips for both signs of $\mu$ and various
$\tan\beta$. Results at $\tan\beta=3.0$ are ruled out by the bound on
$m_h$ set by LEP2, but we nonetheless report them for completeness.}
\label{fig:Coann2}
\end{center}
\end{figure}

We provide in this section some results concerning the branching
ratios of LFV in a particular region of the mSUGRA parameter space,
namely the one where {\em coannihilations} between the lightest
neutralino and the lightest stau (the next-to-lightest SUSY particle)
concur in reducing the neutralino relic density within the current
observational cold dark matter content of the 
universe. In Fig.~\ref{fig:CoannTB30} we study the 
${\rm BR}(\mu\rightarrow e\gamma)$ 
(\emph{a}) and ${\rm BR}(\tau\rightarrow \mu\gamma)$
(\emph{b}) as a function of the neutralino mass $m_\chi$ along the
``coannihilation corridor'' at fixed $\tan\beta=30.0$ for three different
values of $M_R=10^9~\GeV,\ 10^{11}~\GeV$ and $10^{13}~\GeV$. We recall
that in mSUGRA $m_\chi\approx0.4M_{1/2}$. Notice that for sufficiently
low values of $M_R$ the computed ${\rm BR}(\mu\rightarrow e\gamma)$ is
always {\em below} the current experimental bounds, while putative
lower bounds on $m_{\chi}$ can be drawn for larger $M_R$. In
the case of ${\rm BR}(\tau\rightarrow \mu\gamma)$ we always get
results far below (at least two orders of magnitude) the current
experimental bound. As was natural to expect, and as is shown in
Fig.~\ref{fig:Coann2} (\emph{a}), lowering $\tan\beta$ yields a
quadratic suppression in the LFV Branching ratios, which appear to be
in all cases well below the current experimental sensitivity. Finally,
in Fig.~\ref{fig:Coann2} (\emph{b}) we summarize our results showing
the $M_{1/2}$ ranges dictated by cosmological and phenomenological
requirements. These constraints, leading to smaller allowed $M_{1/2}$
for $\mu>0$ indicate that this last case is favored to produce larger
LFV decay rates, which may be in the range of sensitivity
of the next generation experiments.

\subsection{A Candidate for the Effective SUSY Mass $m_{\sss S}$}
\indent
The branching ratio for the processes $l_j\ra l_i\gamma$
is often quoted in the literature in the mass insertion
and leading log approximations as 
\begin{equation}
{\rm BR}(l_j\ra l_i\gamma)\simeq
\frac{\alpha^3}{G_{\sss F}^2} \frac{\abs{\left(m_{\s {\sss L}}^2(LL)\right)_{ji}}^2}{m_{\sss S}^8}
\tan^2\beta\nn, \label{eq:min} 
\end{equation}
where $m_{\sss S}$ is a {\em typical mass} of superparticles,
$\alpha\simeq1/128$ and $\left(m_{\s {\sss L}}^2(LL)\right)_{ji}$ 
is given by Eq.~(\ref{eq:leadinglog}). The problem with this formula
is that there is no prescription for $m_{\sss S}$ in terms of the
fundamental SUSY and soft breaking parameters of the theory, and the
dependence on $m_{\sss S}$ is so strong that the predictions
for ${\rm BR}(l_j\ra l_i\gamma)$ depend drastically on 
what one uses for $m_{\sss S}$: $m_{\sss S}=m_0$, or
$m_{\sss S}=m_{\s\nu}$, or $m_{\sss S}=\sqrt{\mu M_{1/2}}$. The 
three indicated choices will give
completely different predictions for the branching ratios of 
interest. Notice that
if we underestimate $m_{\sss S}$ by just a factor of two we are
overestimating the branching ratios by more than two orders 
of magnitude. 

We have found that the expression~(\ref{eq:min}) 
with $\left(m_{\s {\sss L}}^2(LL)\right)_{ji}$ given by
Eq.~(\ref{eq:leadinglog}) represents an excellent approximation
to the exact RG result if for $m_{\sss S}$ one uses 
\begin{equation} 
m_{\sss S}^8\simeq 0.5\ m_0^2\ M_{1/2}^2\ \left(m_0^2\ +\ 0.6\ M_{1/2}^2\right)^2 \nn.
\label{eq:ms2}
\end{equation}

\begin{figure}[!h]
\begin{center}
\includegraphics[scale=0.5,angle=-90]{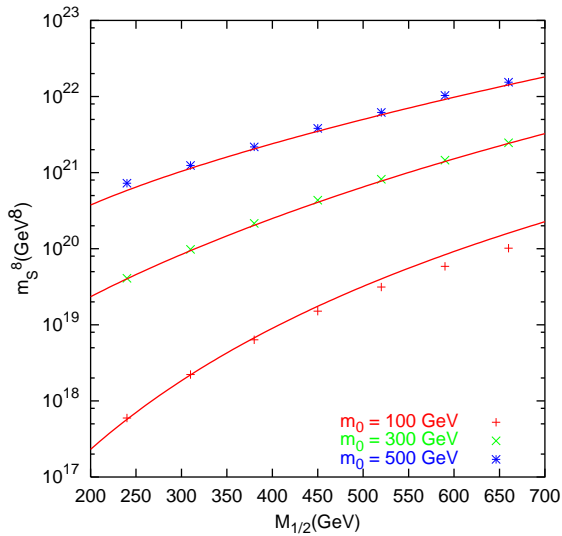} 
\caption{ \em\small A fit to $m_{\sss S}^8$. The lines 
correspond to Eq.~(\ref{eq:ms2}), while the points to the 
value extracted from the full RGE numerical results and Eq.~(\ref{eq:min}). In 
this plot we use $M_R=10^{14}~\GeV$, $\tan\beta=3$ 
and $A_0=0$.}
\label{fig:ms}
\end{center}
\end{figure}

In Fig.~\ref{fig:ms} we compare the predictions for $m_{\sss S}^8$
obtained from (\ref{eq:min}) and the exact RG result for 
${\rm BR}(l_j\ra l_i\gamma)$ with that given
by Eq.~(\ref{eq:ms2}). In general, we find that our fit formula
(\ref{eq:ms2}) slightly underestimates $m_{\sss S}$ at low $M_{1/2}$
and large $m_0$, and somewhat overestimates it at large
$M_{1/2}$ and small $m_0$. The deviations are, however, always 
relatively small, and the overall dependence of $m_{\sss S}^8$
on $m_0$ and $M_{1/2}$ is everywhere reproduced rather accurately.

\section{Conclusions}
\label{sec:conclusions}
\indent
We have shown that the off-diagonal elements of the slepton mass
matrix obtained by RG running deviate significantly -- at low $m_0$ and
large $M_{1/2}$ -- from the values  given by the leading logarithmic
approximation. The exact result is always larger than the approximated
one and this difference increases for larger $M_R$. As a consequence,
in the leading log approximation the predictions for $\mu\ra e\gamma$
and $\tau\ra \mu\gamma$ can be underestimated by a factor $\sim10$. We
pointed out that, even when this factor is smaller, it is relevant
if one wants to use ${\rm BR}(\mu\ra e\gamma)$ and 
${\rm BR}(\tau\ra \mu\gamma)$
to constrain the SUSY parameter space. These branching ratios 
were found to depend on the sign of $A_0$ and of $\mu$ as well.

We also studied the behavior of ${\rm BR}(\mu\ra e\gamma)$ and 
${\rm BR}(\tau\ra\mu\gamma)$ along the cosmologically determined 
``coannihilation strips''. Our results show that
the present experimental bound on $\mu\ra e\gamma$ can exclude only
regions with large $M_R$ ($M_R\geq 10^{13}~\GeV$), while 
$\tau\ra \mu\gamma$ is always allowed. Finally, we proposed a simple
parametrization in terms of $m_0$ and
$M_{1/2}$ for the effective SUSY mass parameter $m_{\sss S}$  
which enters into the simple expression for ${\rm BR}(l_j\ra l_i\gamma)$
derived in the leading log and mass insertion 
approximations (see Eqs.~(\ref{eq:leadinglog})-(\ref{eq:ms2})). This 
allows to reproduce 
the exact RG results for ${\rm BR}(l_j\ra l_i\gamma)$ with 
high precision.

\section*{Acknowledgments}
\indent 
One of us (Y.T.) wishes to thank 
H.~B.~Nielsen and G.~Senjanovi{\'c} 
for useful discussions. This work was supported in part 
by the EC network HPRN-CT-2000-00152 (Y.T.), 
and the Italian MIUR and INFN under the programs 
``Fenomenologia delle Interazioni Fondamentali'' 
and ``Fisica Astroparticellare'' (S.T.P., S.P. and C.E.Y.).

\appendix
\section{Renormalization Group Equations}
\label{sec:appA} 
\indent
{}From the GUT scale $M_{\rm\sss GUT}$ down to the 
universal $M_{\rm\sss R}$ scale we use the one-loop 
MSSM renormalization group equation~\cite{RGE}\footnote{We adopt 
S.~P.~Martin and M.~T.~Vaughn convention~\cite{RGE}, $\ie$, the sign of 
the terms proportional to the gaugino masses in the trilinear 
parameters, $A$, are different from those 
used in~\cite{MSSMRN}.} of the 
gauge coupling
constants, the three gaugino mass parameters, the Yukawa coupling 
constant matrices, $Y_{u}$, $Y_{d}$, $Y_{e}$, $Y_{\nu}$,
and the right-handed neutrino mass matrix, $M_{\sss R}$, and 
the soft supersymmetry breaking mass terms and trilinear 
parameters, $A$.

The gauge coupling constants (especially $g_2$ running
effect) and the gaugino mass terms (as well $M_2$ running effect), 
play important r{\^o}le in the calculation of 
the lepton flavor violation processes, therefore 
a part of the two-loop $\beta$ functions are adopted in our computations. 
We present at first 
the non-supersymetric part of the renormalization from the GUT 
scale $M_{\rm\sss GUT}$ to the universal $M_{\sss R}$ scale:
\begin{eqnarray}
16\pi^2 \frac{d}{d t} ~g_1 &\!=\!& \frac{33}{5}\, g_1^3 
+ \frac{g_1^3}{16\pi^2} \left(\frac{199}{25} g_1^2 + \frac{27}{5} g^2_2
+ \frac{88}{5} g_3^2\right) \nn, \\
16\pi^2 \frac{d}{d t} ~g_2 &\!=\!& \, g_2^3 + 
\frac{g_2^3}{16\pi^2} \left(\frac{9}{5} g_1^2 + 25 g^2_2
+ 24 g_3^2\right) \nn, \\
16\pi^2 \frac{d}{d t} ~g_3 &\!=\!& -3 \, g_3^3 
+ \frac{g_3^3}{16\pi^2} \left(\frac{1}{5} g_1^2 + 9 g^2_2
+ 14 g_3^2\right) \nn, \\
16\pi^2 \frac{d}{d t} Y_{u_{ij}} &\!=\!& 
\left \{ -\frac{13}{15} g_1^2 - 3 g_2^2 -\frac{16}{3} g_3^2 
  + 3 \,{\rm Tr} ( Y_u Y_u^{\dagger} )
  +   {\rm Tr} ( Y_\nu Y_\nu^{\dagger} ) \right \} Y_{u_{ij}}\nonumber \\
&& \; + 3 \, ( Y_u Y_u^{\dagger} Y_u)_{ij} 
      + ( Y_u Y_{d}^{\dagger} Y_{d})_{ij}\nn, \\
%
16\pi^2 \frac{d}{d t} Y_{d_{ij}} &\!=\!& 
\left \{ -\frac{7}{15} g_1^2 - 3 g_2^2 -\frac{16}{3} g_3^2 
  + 3 \,{\rm Tr} ( Y_d Y_d^{\dagger} )
  +   {\rm Tr} ( Y_e Y_e^{\dagger} ) \right \} Y_{d_{ij}} \nonumber \\
&& 
+ \; 3 \, ( Y_d Y_d^{\dagger} Y_d)_{ij} 
+ ( Y_d Y_{u}^{\dagger} Y_{u})_{ij}\nn, \\
16\pi^2 \frac{d}{d t} Y_{e_{ij}} &\!=\!& 
\left \{ -\frac{9}{5} g_1^2 - 3 g_2^2 
  + 3 \,{\rm Tr} ( Y_d Y_d^{\dagger})
  +   {\rm Tr} ( Y_e Y_e^{\dagger}) 
\right \} Y_{e_{ij}}\nonumber \\
&& 
+ 3 \, \left( Y_e Y_e^{\dagger} Y_e \right)_{ij} + 
\left( Y_e Y_{\nu}^{\dagger} Y_{\nu} \right)_{ij}\nn, \\
16\pi^2 \frac{d}{d t} Y_{\nu_{ij}}&\!=\!&
\left \{ -\frac{3}{5} g_1^2 - 3 g_2^2 
+ 3 \,{\rm Tr} \left( Y_{u} Y_{u}^{\dagger} \right)
+   {\rm Tr} \left( Y_{\nu} Y_{\nu}^{\dagger} \right) 
\right \} Y_{\nu_{ij}}  \nonumber \\
&& 
+ 3 \, \left( Y_{\nu} Y_{\nu}^{\dagger} Y_{\nu} \right)_{ij}
+ \left( Y_{\nu} Y_{e}^{\dagger} Y_{e} \right)_{ij}\nn, \\
16\pi^2 \frac{d}{d t} M_{R_{ij}}&\!=\!&
 2 \left( M_R Y^\dagger_\nu Y_\nu \right)_{ij} 
+ 2 \left( Y_\nu Y^\dagger_\nu M_R \right)_{ij} \nn.
\end{eqnarray}

Here are the renormalization group equations of the gaugino mass terms, 
the soft mass terms and the trilinear parameters as well the up-type
and down-type Higgs mass terms, respectively. Note that we take into
account the two-loop $\beta$ function accuracy only for the
gaugino mass terms:
\begin{eqnarray}
16\pi^2 \frac{d}{d t} M_1 &\!=\!& \frac{66}{5} \, g_1^2 M_1 +
 \frac{2 g_1^2}{16\pi^2} \left\{ \frac{199}{5} g^2_1 \left( 2 M_1  
\right) \right. \nonumber \\
&& \left. + \frac{27}{5} g_2^2 \left( M_1 + M_2 \right) 
+ \frac{88}{5} g_3^2 \left( M_1 + M_3 \right)\right\} \nn, \\
16\pi^2 \frac{d}{d t} M_2 &\!=\!& 2 \, g_2^2 M_2 + 
\frac{2 g_2^2}{16\pi^2} \left\{ \frac{9}{5} g^2_1 \left(M_1 + M_2 \right)
\right. \nonumber \\
&& \left. + 25 g_2^2 \left( 2 M_2\right) + 24 g_3^2 
\left( M_2 + M_3 \right) \right\} \nn, \\
16\pi^2 \frac{d}{d t} M_3 &\!=\!& -6\, g_3^2 M_3 + 
\frac{2 g_3^2}{16\pi^2} \left\{ \frac{11}{5} g^2_1 \left(M_1 + M_3 \right)
\right. \nonumber \\
&& \left. + 9 g_2^2 \left( M_2 + M_3 \right) + 14 g_3^2 
\left( 2 M_3 \right) \right\} \nn, \\
16\pi^2 \frac{d}{d t} \left( m^2_{\widetilde{\sss Q}} \right)_{ij} 
&\!=\!&
- \left( \frac{2}{15} g_1^2 \left| M_1 \right|^2 
+ 6 g_2^2 \left| M_2 \right|^2 + \frac{32}{3} g_3^2 \left| M_3 
\right|^2\right) \delta_{ij}
+ \frac{1}{5} g_1^2~S~\delta_{ij} \nonumber \\
&& 
+ \left( m^2_{\widetilde{\sss Q}} Y_u^{\dagger} Y_u 
+ m^2_{\widetilde{\sss Q}} Y_d^{\dagger} Y_d 
+ Y_u^{\dagger} Y_u m^2_{\widetilde{\sss Q}} 
+ Y_d^{\dagger} Y_d  m^2_{\widetilde{\sss Q}} \right)_{ij} \nonumber \\
&&
+ 2 \left( Y_u^{\dagger} m^2_{\widetilde u} Y_u
           + {m}^2_{H_u} Y_u^{\dagger} Y_u
+ A_u^{\dagger} A_u \right)_{ij} \nonumber \\
&&
+ 2 \left( Y_d^{\dagger} m^2_{\widetilde {d}} Y_{d}
+ {m}^2_{H_d} Y_{d}^{\dagger} Y_{d}
+ A_{d}^{\dagger} A_{d} \right)_{ij}\nn, \\
%
16\pi^2 \frac{d}{d t} \left( m^2_{\widetilde{u}} \right)_{ij} 
&\!=\!&
- \left( \frac{32}{15} g_1^2 \left| M_1 \right|^2 
+ \frac{32}{3} g_3^2 \left| M_3 \right|^2\right) \delta_{ij}
- \frac{4}{5} g_1^2~S~\delta_{ij} \nonumber \\
&& 
+ 2 \left( m^2_{\widetilde u} 
Y_u Y_u^{\dagger} + Y_u Y_u^{\dagger} m^2_{\widetilde u} \right)_{ij} 
\nonumber \\
&&
+ 4 \left( Y_u m^2_{\widetilde{\sss Q}} Y_u^{\dagger} + {m}^2_{H_u} 
Y_u Y_u^{\dagger} + A_u A_u^{\dagger} \right)_{ij}\nn, \\
16\pi^2 \frac{d}{d t} \left( m^2_{\widetilde{d}} \right)_{ij} 
&\!=\!&
- \left( \frac{8}{15} g_1^2 \left| M_1 \right|^2 
+ \frac{32}{3} g_3^2 \left| M_3 \right|^2\right) \delta_{ij}
+ \frac{2}{5} g_1^2~S~\delta_{ij} \nonumber \\
&& 
+ 2 \left( m^2_{\widetilde d} 
Y_d Y_d^{\dagger} + Y_d Y_d^{\dagger} m^2_{\widetilde d} \right)_{ij} 
\nonumber \\
&&
+ 4 \left( Y_d m^2_{\widetilde{\sss Q}} Y_d^{\dagger} + {m}^2_{H_d} 
Y_d Y_d^{\dagger} + A_d A_d^{\dagger} \right)_{ij}\nn, \\
%
%
16\pi^2 \frac{d}{d t} \left( m^2_{\widetilde{\sss L}} 
\right)_{ij}&\!=\!&   -\left( \frac{6}{5} g_1^2 \left| M_1 \right|^2 
+ 6 g_2^2 \left| M_2 \right|^2 \right) \delta_{ij}
-\frac{3}{5} g_1^2~S~\delta_{ij} \nonumber \\
&& 
+ \left( m^2_{\widetilde{\sss L}} Y_e^{\dagger} Y_e 
+ m^2_{\widetilde{\sss L}} Y_\nu^{\dagger} Y_\nu 
+ Y_e^{\dagger} Y_e m^2_{\widetilde{\sss L}} 
+ Y_\nu^{\dagger} Y_\nu  m^2_{\widetilde{\sss L}} \right)_{ij} \nonumber \\
&&
+ 2 \left( Y_e^{\dagger} m^2_{\widetilde e} Y_e
           +{m}^2_{H_d} Y_e^{\dagger} Y_e
+ A_e^{\dagger} A_e \right)_{ij} \nonumber \\
&&
+ 2 \left( Y_{\nu}^{\dagger} m^2_{\widetilde {\sss\nu}} Y_{\nu}
+ {m}^2_{H_u} Y_{\nu}^{\dagger} Y_{\nu}
+ A_{\nu}^{\dagger} A_{\nu} \right)_{ij}\nn, \\
16\pi^2 \frac{d}{d t} \left( m^2_{\widetilde{e}} \right)_{ij}&\!=\!& 
- \frac{24}{5} g_1^2 \left| M_1 \right|^2 \delta_{ij}
+ \frac{6}{5} g_1^2~S~\delta_{ij} + 2 \left( m^2_{\widetilde e} 
Y_e Y_e^{\dagger} + Y_e Y_e^{\dagger} m^2_{\widetilde e} \right)_{ij} 
\nonumber \\
&&
+ 4 \left( Y_e m^2_{\sss\widetilde L} Y_e^{\dagger} + {m}^2_{H_d} 
Y_e Y_e^{\dagger} + A_e A_e^{\dagger}\right)_{ij}\nn, \\
16\pi^2 \frac{d}{d t} \left( m^2_{\widetilde{\nu}} \right)_{ij} &\!=\!& 
2 \left( m^2_{\widetilde \nu} Y_{\nu} Y_{\nu}^{\dagger} 
+ Y_{\nu} Y_{\nu}^{\dagger} m^2_{\widetilde \nu} \right)_{ij} 
+ 4 \left( Y_{\nu} m^2_{\widetilde L} Y_{\nu}^{\dagger}
+ {m}^2_{H_u} Y_{\nu} Y_{\nu}^{\dagger}
+ A_{\nu} A_{\nu}^{\dagger}\right)_{ij} \nn, \\
16\pi^2 \frac{d}{d t} A_{e_{ij}}&\!=\!&  
 \left\{ -\frac{9}{5} g_1^2 -3 g_2^2
+ 3 {\rm Tr} ( Y_d^{\dagger} Y_d )
+   {\rm Tr} ( Y_e^{\dagger} Y_e ) \right \} A_{e_{ij}} \nonumber \\
&&
+ 2 \left\{
\frac{9}{5} g_1^2 M_1 + 3 g_2^2 M_2 
+ 3 {\rm Tr} ( Y_d^{\dagger} A_d)
+   {\rm Tr} ( Y_e^{\dagger} A_e) \right \} Y_{e_{ij}} \nonumber \\
&&
+ 4 \left( Y_e Y_e^{\dagger} A_e \right)_{ij} 
+ 5 \left( A_e Y_e^{\dagger} Y_e \right)_{ij} 
+ 2 \left( Y_e Y_{\nu}^{\dagger} A_{\nu} \right)_{ij} 
+  \left( A_e Y_{\nu}^{\dagger} Y_{\nu} \right)_{ij} \nn, \\
16\pi^2 \frac{d}{d t} A_{\nu_{ij}}&\!=\!&  
\left\{ -\frac{3}{5} g_1^2 -3 g_2^2 
+ 3 {\rm Tr} ( Y_u^{\dagger} Y_u)
+  {\rm Tr} ( Y_{\nu}^{\dagger} Y_{\nu}) \right \} A_{\nu_{ij}} \nonumber \\
&&
+ 2 \left\{ \frac{3}{5} g_1^2 M_1 + 3 g_2^2 M_2 
+ 3 {\rm Tr} ( Y_u^{\dagger} A_u)
+   {\rm Tr} ( Y_{\nu}^{\dagger} A_{\nu}) \right \} Y_{\nu_{ij}} 
\nonumber \\
&&
+ 4 ( Y_{\nu} Y_{\nu}^{\dagger} A_{\nu})_{ij}
+ 5 ( A_{\nu} Y_{\nu}^{\dagger} Y_{\nu})_{ij} 
+ 2 ( Y_{\nu} Y_e^{\dagger} A_e)_{ij}
+ ( A_{\nu} Y_e^{\dagger} Y_e)_{ij} \nn, \\
16\pi^2 \frac{d}{d t} A_{u_{ij}}&=&
\left\{ -\frac{13}{15} g_1^2 - 3 g_2^2 - \frac{16}{3} g_3^2 + 3 {\rm Tr} 
(Y_u^{\dagger} Y_u) + {\rm Tr}( Y_{\nu}^{\dagger} Y_{\nu}) 
\right \} A_{u_{ij}} 
\nonumber \\
&& 
+ 2 \left\{ \frac{13}{15} g_1^2 M_1 + 3 g_2^2 M_2 +\frac{16}{3} g_3^2 M_3
+ 3 {\rm Tr}(Y_u^{\dagger} A_u)
+ {\rm Tr}( Y_{\nu}^{\dagger} A_{\nu}) \right \} Y_{u_{ij}}
\nonumber \\
&& 
+ 4 ( Y_u Y_u^{\dagger} A_u)_{ij} 
+ 5 ( A_u Y_u^{\dagger} Y_u)_{ij} 
+ 2 ( Y_u Y_d^{\dagger} A_d)_{ij} 
+ ( A_u Y_d^{\dagger} Y_d)_{ij} \nn,\\
16 \pi^2 \frac{d}{d t} A_{d_{ij}}&=&
\left\{
-\frac{7}{15} g_1^2 - 3 g_2^2  -\frac{16}{3} g_3^2 
+ 3 {\rm Tr} ( Y_d^{\dagger} Y_d)
+ {\rm Tr}( Y_{e}^{\dagger} Y_{e}) \right \} A_{d_{ij}}
\nonumber \\
&& + 2 \left\{ \frac{7}{15} g_1^2 M_1 + 3 g_2^2 M_2 + 
\frac{16}{3} g_3^2 M_3 + 3 {\rm Tr}( Y_d^{\dagger} A_d) 
+ {\rm Tr} ( Y_{e}^{\dagger} A_{e}) \right \} Y_{d_{ij}}
\nonumber \\ 
&& 
+ 4 ( Y_d Y_d^{\dagger} A_d)_{ij} 
+ 5 ( A_d Y_d^{\dagger} Y_d)_{ij} 
+ 2 ( Y_d Y_u^{\dagger} A_u)_{ij} 
+ ( A_d Y_u^{\dagger} Y_u)_{ij} \nn, \\
%
{16 \pi^2} \frac{d}{d t} (m^2_{H_u}) &=&  -\left(\frac{6}{5} 
g^{2}_{1} \abs{M_{1}}^{2} + 6 g^{2}_{2} \abs{M_{2}}^{2}\right)
+ \frac{3}{5} g_1^2 S
\nonumber\\
&& 
+ 6\, {\rm {\rm Tr}}\left(
m^2_{\widetilde Q} Y^{\dagger}_{u} Y_{u} + Y^{\dagger}_{u}
( m^2_{\widetilde u} + m^2_{H_u} ) Y_{u} + A^{\dagger}_{u} A_{u}
\right) \nonumber\\
&&  + 2\, {\rm Tr} \left( m^2_{\widetilde{\sss L}} Y^{\dagger}_{\nu} Y_{\nu}
+ Y^{\dagger}_{\nu} ( m^2_{\widetilde \nu} + m^2_{H_u}) Y_{\nu} 
+ A^{\dagger}_{\nu} A_{\nu} \right) \nn,\\
{16 \pi^2} \frac{d}{d t} (m^2_{H_d}) &=& -\left( \frac{6}{5} 
g^{2}_{1} \abs{M_{1}}^{2} + 6 g^{2}_{2} \abs{M_{2}}^{2}\right) 
- \frac{3}{5} g_1^2 S \nonumber\\
&& 
+ 6\, {\rm Tr}\left( m^2_{\widetilde Q} Y^{\dagger}_{d} Y_{d} 
+ Y^{\dagger}_{d} 
( m^2_{\widetilde d} + m^2_{H_d}) Y_{d} + A^{\dagger}_{d} A_{d}\right)
\nonumber\\
&& + 2\, {\rm Tr} \left( m^2_{\widetilde{\sss L}} Y^{\dagger}_{e} Y_{e} 
+ Y^{\dagger}_{e} ( m^2_{\widetilde e} + m^2_{H_d}) Y_{e} 
+ A_{e}^{\dagger} A_{e} \right) \nn,
\end{eqnarray}
where
\begin{equation}
S = {\rm Tr} (m_{\widetilde Q}^2 + m_{\widetilde d}^2 - 2 m_{\widetilde u}^2
- m_{\widetilde L}^2 + m_{\widetilde e}^2 ) - {m}^2_{H_d} 
+ {m}^2_{H_u} \nn,
\end{equation}
and we have used the GUT convention for the 
$U(1)$ gauge coupling constant, $g_1$, and $t=\ln\mu$ where
$\mu$ is denoted as the renormalization point.

\section{Notations in the MSSM}
\label{sec:appB}
\indent

In this appendix we set the notations and conventions 
for the computation of LFV processes. Note that
we follow the Appendix B of ref.~\cite{MSSMRN}.
 
The chargino mass matrix has the following form:
 \begin{equation}
   -\mathscr{L}_m =
   \left ( \overline{\widetilde W^-_{\sss R}}~ 
     \overline{\widetilde H^-_{2{\sss R}}} \right )
   M_{\sss C}
   \left ( \begin{array}{c}
       \widetilde W^-_{\sss L}    \\ 
        \widetilde H_{1{\sss L}}^-
     \end{array}
\right) + h.c.\nn,
 \end{equation}
where
\begin{equation}
  \label{eq:charginomassmatrix}
  M_{\sss C} = \left (  \begin{array}{cc}
       M_2 & \sqrt{2} m_W \cos \beta \\
       \sqrt{2}m_W \sin \beta &  \mu
     \end{array} \right) \nn. 
\end{equation}
This matrix can be diagonalized
by two $2 \times 2$ real orthogonal matrices $O_{\sss L}$ 
and $O_{\sss R}$ according to:
\begin{equation}
     O_{\sss R} M_{\sss C} O_{\sss L}^{\sf T} ={\rm diagonal}\nn.
\end{equation}
If we define
\begin{equation}
  \left( \begin{array}{c}
      \widetilde \chi^-_{1 {\sss L}} \\
      \widetilde \chi^-_{2{\sss L}}
    \end{array}\right)
  =O_{\sss L}   \left( \begin{array}{c}
      \widetilde W^-_{\sss L}  \\
      \widetilde H^-_{1{\sss L}}
    \end{array}\right) \nn,
  \hspace{1cm}
  \left( \begin{array}{c}
      \widetilde \chi^-_{1 {\sss R}} \\
      \widetilde \chi^-_{2{\sss R}}
    \end{array} \right)
  = O_{\sss R}   \left( \begin{array}{c}
      \widetilde W^-_{\sss R}  \\
      \widetilde H^-_{2{\sss R}}
    \end{array} \right) \nn,
\end{equation}
then
\begin{equation}
    {\widetilde \chi}^-_{\sss A}  = {\widetilde \chi}^-_{{\sss A L}} 
                           + {\widetilde \chi}^-_{{\sss A R}}
\end{equation} %
forms a Dirac fermion with mass
$M_{{\widetilde \chi}^-_{\sss A}}$.

The mass matrix of the neutralino sector is given by
\begin{equation}
  -\mathscr{L}_m =
  \frac{1}{2} \left( {\widetilde B}_L {\widetilde W}^0_L {\widetilde H}^0_{1L} 
    {\widetilde H}^0_{2L} \right) M_N 
  \left(   
    \begin{array}{c}
      {\widetilde B}_L  \\
      {\widetilde W}^0_L \\
      {\widetilde H}^0_{1L} \\
      {\widetilde H}^0_{2L}
    \end{array}
  \right)  + h.c. \nn,
\end{equation}
where
\begin{equation}
   M_{\sss N}=
   \left(
   \begin{array}{cccc}
     M_1    & 0 & -m_Z\sin\theta_W\cos\beta & m_Z\sin\theta_W\sin\beta \\
     0 & M_2 & m_Z\cos\theta_W\cos\beta & -m_Z\cos\theta_W\sin\beta \\
     -m_Z\sin\theta_W\cos\beta & m_Z\cos\theta_W\cos\beta & 0 & -\mu
     \\
     m_Z\sin\theta_W\sin\beta & -m_Z\cos\theta_W\sin\beta & -\mu & 0
   \end{array}            \right) \nn.
\end{equation}
We can diagonalize $M_{\sss N}$ with a real 
orthogonal matrix $O_{\sss N}$:
\begin{equation}
    O_{\sss N} M_{\sss N} O_{\sss N}^{\sf T} = {\rm diagonal} \nn.
\end{equation}
The mass eigenstates are given by
\begin{equation}
\widetilde \chi^0_{\sss BL} =(O_N)_{BC} \widetilde X^0_{\sss CL} \nn
           (B,C=1, \cdots ,4) \nn\nn; \nn
   \widetilde X^0_{CL} = ( \widetilde B_{\sss L}, \widetilde W^0_{\sss L}, 
\widetilde H^0_{1{\sss L}},
   \widetilde H^0_{2{\sss L}}) \nn.
\end{equation}
We have four Majorana spinors
\begin{equation}
   \widetilde \chi^0_{\sss B} = \widetilde \chi^0_{\sss B L} 
+ \widetilde \chi^0_{\sss B R}\nn, \hspace{1cm}(B=1, \cdots ,4)
\end{equation}
with masses $M_{\widetilde \chi^0_{\sss B}}$.


The slepton mass matrix can be cast in the following form:
 \begin{equation}
   -\mathscr{L}_s =
 \left ( \widetilde e^{\dagger}_{\sss L}, 
\widetilde e^{\dagger}_{\sss R} \right )
     \left ( \begin{array}{cc}
               m_{\sss L}^2    & m_{\sss LR}^{2 {\dagger}} \\
               m_{\sss LR}^{2} & m_{\sss R}^2
             \end{array}                  \right )
     \left (  \begin{array}{c}
              \widetilde e_{\sss L} \\ \widetilde e_{\sss R}
              \end{array} \right )\nn,
\end{equation}
with
\begin{eqnarray}
\left(m_{\sss L}^2\right)_{ij} &=& \left(m_{\widetilde L}^2\right)_{ij} 
+ m_{e_i}^2 \delta_{ij} - m_Z^2 \cos2\beta \left(\frac{1}{2}+ 
\sin^2 \theta_{\sss W}\right) \delta_{ij}\nn, \\
\left(m_{\sss R}^2\right)_{ij} &=& \left( m_{\widetilde e}^2 \right)_{ij}
+ m_{e_i}^2\delta_{ij} - m_Z^2 \cos 2 \beta\sin^2 \theta_{\sss W}  \delta_{ij}\nn, \\
\left(m_{\sss LR}^2\right)_{ij} &=& \frac{v \cos\beta}{\sqrt{2}} A_{e ij} -
 m_{e_i} \mu \tan\beta \delta_{ij}\nn.
\end{eqnarray}
We diagonalize the slepton mass matrix, $M^2_{\sss L}$, 
by a $6 \times 6$ real orthogonal matrix $U^l$ as
\begin{equation}
    U^l M^2_{\sss L} U^{l {\sf T}} = {\rm diag.}\left(m^2_{{\widetilde l}_{1}}, 
\cdots, m^2_{{\widetilde l}_{ 6}}\right) \nn,
\end{equation}
A mass eigenstate is then written as
\begin{equation}
   {\widetilde l}_{\sss Y} = U^l_{{\sss Y},i} {\widetilde l}_{{\sss L} i} 
+ U^l_{{\sss Y}, i+3} {\widetilde l}_{{\sss R}i}\nn, \hspace{1cm} (Y=1, \cdots, 6) \nn.
\end{equation}
Since there are no right-handed sneutrinos at low energies, the sneutrino mass matrix is 
\begin{equation}
(M_{\s\nu}^2)_{ij}=\left(m_{\widetilde L}^2\right)_{ij} 
+ \frac 12 m_Z^2 \cos2\beta \delta_{ij}\nn. \\
\end{equation} 
The diagonalization is carried out by a $3\times 3$ orthogonal matrix $U^\nu$
\begin{equation}
U^\nu M_{\s\nu}^2U^{\nu{\sf T}}={\rm diag.}
\end{equation}
and the mass eigenstates read
\begin{equation}
\widetilde\nu_X=U^\nu_{X,i} \widetilde\nu_{Li}, \quad X=1,2,3.
\end{equation}

The amplitudes $A_{\sss L,R}$ for the processes $l_j\to l_i\gamma$ are
given by the sum of the chargino and neutralino contributions:
\begin{eqnarray}
\label{eq:amp}
 A_{\sss L,R} \!\!&=&\!\!  A^{(c)}_{\sss L,R} + A^{(n)}_{\sss L,R} \nn,\\
\label{eq:cgino}
A^{(c)}_{\sss L}  \!\!&=&\!\! -\frac{1}{32\pi^{2}}\sum_{A,X}
\frac{1}{m_{\widetilde{\nu}_{X}}^{2}}
\left[ C_{j{\sss A}{\sss X}}^{{\sss L}(l)} C_{i{\sss A}{\sss X}}^{{\sss L}(l)*} 
\frac{1}{6(1-x_{{\sss A}{\sss X}})^{4}}(2+3x_{{\sss A}{\sss X}}
-6x^{2}_{{\sss A}{\sss X}}
+x^{3}_{{\sss A}{\sss X}}+6x_{{\sss A}{\sss X}}\ln x_{{\sss A}{\sss X}})
\right.\nonumber\\
&& \left.\hspace{2.0 cm}+
C_{j{\sss A}{\sss X}}^{{\sss L}(l)}C_{i{\sss A}{\sss X}}^{{\sss R}(l)*}
\frac{M_{\widetilde{\chi}_{{\sss A}}^{-}}}{m_{j}}
\frac{1}{(1-x_{{\sss A}{\sss X}})^{3}}(-3+4x_{{\sss A}{\sss X}}
-x^{2}_{{\sss A}{\sss X}}-2\ln x_{{\sss A}{\sss X}})\right] \hspace{1mm},\\
\label{eq:nulino}
A^{(n)}_{\sss L}  \!\!&=&\!\! \phantom{-}\frac{1}{32\pi^{2}}\sum_{B,Y}
\frac{1}{m_{\widetilde{l}_{{\sss Y}}}^{2}}
\left[N_{j{\sss B}{\sss Y}}^{{\sss L}(l)}N_{i{\sss B}{\sss Y}}^{{\sss L}(l)*}
\frac{1}{6(1-y_{{\sss B}{\sss Y}})^{4}}(1-6y_{{\sss B}{\sss Y}}
+3y^{2}_{{\sss B}{\sss Y}} +2y^{3}_{{\sss B}{\sss Y}}
-6y^{2}_{{\sss B}{\sss Y}} \ln y_{{\sss B}{\sss Y}})\right.\nonumber\\
&&\left.\hspace{2.0cm}+
N_{j{\sss B}{\sss Y}}^{{\sss L}(l)}N_{i{\sss B}{\sss Y}}^{{\sss R}(l)*} 
\frac{M_{\widetilde{\chi}_{{\sss B}}^{0}}}{m_{i}} 
\frac{1}{(1-y_{{\sss B}{\sss Y}})^{3}} 
(1+y^{2}_{{\sss B}{\sss Y}}+2y_{{\sss B}{\sss Y}}
\ln y_{{\sss B}{\sss Y}})\right]\ ,\\
A^{(c,n)}_{\sss R}  \!\!&=&\!\! A^{(c,n)}_{\sss L}\big|_{\sss L\leftrightarrow R} \nn.
\end{eqnarray}
where the dimensionless parameters $x_{\sss AX}$ and $y_{\sss BY}$ are defined as
\begin{equation}
x_{{\sss A}{\sss X}} =
\frac{M_{\widetilde{\chi}^{-}_{\sss A}}^{2}}{m_{\widetilde{\nu}_{\sss X}}^{2}} 
\nn,\nn\nn\nn
y_{{\sss B}{\sss Y}}=\frac{M_{\widetilde{\chi}^{0}_{\sss B}}^{2}}
{m_{\widetilde{l}_{\sss Y}}^{2}} \nn,
\end{equation}
and the coefficients in Eqs.~(\ref{eq:cgino}) and (\ref{eq:nulino}) are given by
\begin{eqnarray}
C^{{\sss L}(l)}_{i{\sss AX}}& = & g_2\frac{m_{l_i}}{\sqrt{2}m_W\cos\beta}
(O_{\sss L})_{{\sss A}2}U^{\nu}_{{\sss X},i} \nn,\\
C^{{\sss R}(l)}_{i{\sss AX}}& =& -g_2(O_{\sss R})_{{\sss A}1} U^{\nu}_{{\sss X},i} 
\nn,\\
N^{{\sss L}(l)}_{i{\sss BY}} &=& -\frac{g_2}{\sqrt{2}} 
\left\{\frac{m_{l_i}}{m_W\cos\beta} (O_{\sss N})_{{\sss B}3} U^{l}_{{\sss Y},i} 
+ 2 (O_{\sss N})_{{\sss B}1} \tan\theta_W U^{l}_{{\sss Y},i+3} \right\} \nn, \\
N^{{\sss R}(l)}_{i{\sss BY}} &=& -\frac{g_2}{\sqrt{2}} 
\left\{[-(O_{\sss N})_{{\sss B}2} 
-(O_{\sss N})_{{\sss B}1} \tan\theta_W] U^l_{{\sss Y},i} 
+ \frac{m_{l_i}}{m_W\cos\beta}(O_{\sss N})_{{\sss B}3} U^l_{{\sss Y},i+3} 
\right\} \nn.
\end{eqnarray}
Using the above mentioned amplitudes, the branching ratio of the decay
$l_j\to l_i\gamma$ is given by
\begin{equation}
  \label{eq:BR}
  {\rm BR}(l_j\to l_i\gamma) = \frac{48 \pi^3\alpha}{G^2_{\sss F}} \left( 
\abs{A_{\sss L}}^2 +  \abs{A_{\sss R}}^2 \right) \nn,
\end{equation}
where $G_{\sss F}$ is the Fermi constant and $\alpha=e^2/4\pi$.


\end{document}